\documentclass[conference]{IEEEtran}
\IEEEoverridecommandlockouts
\usepackage{amsmath,amssymb,amsfonts}
\usepackage{graphicx}
\usepackage{textcomp}
\usepackage{xcolor}

\usepackage{graphicx} 
\usepackage{algorithm}
\usepackage{algorithmicx}
\usepackage[noend]{algpseudocode}
\usepackage{amsthm}
\newtheorem{definition}{Definition}
\newtheorem{theorem}{Theorem}
\newtheorem{proposition}[theorem]{Proposition}

\usepackage{amsmath}
\usepackage{xcolor} 
\usepackage{todonotes}
\usepackage{booktabs}
\usepackage{array}    
\usepackage{url}
\usepackage{colortbl}
\usepackage{hhline}
\usepackage{subfig}
\newtheorem{problem}{Problem}

\newtheoremstyle{proofsketchstyle}
  {}
  {}
  {}
  {}
  {}
  {:}
  {.5em}
  {}
\theoremstyle{proofsketchstyle}

\newtheoremstyle{examplestyle}
  {3pt}
  {3pt}
  {}
  {}
  {\bfseries}
  {.}
  {.5em}
  {}
\theoremstyle{examplestyle}
\newtheorem{example}{Example}

\def\BibTeX{{\rm B\kern-.05em{\sc i\kern-.025em b}\kern-.08em
    T\kern-.1667em\lower.7ex\hbox{E}\kern-.125emX}}
\begin{document}

\title{Learning Dependency Models for Subset Repair}

\author{\IEEEauthorblockN{Haoda Li \IEEEauthorrefmark{2}, 
Jiahui Chen \IEEEauthorrefmark{2}, 
Yu Sun \IEEEauthorrefmark{2},
Shaoxu Song \IEEEauthorrefmark{4},
Haiwei Zhang \IEEEauthorrefmark{2},
Xiaojie Yuan \IEEEauthorrefmark{2}}
\IEEEauthorblockA{\IEEEauthorrefmark{2} \textit{Nankai University}, \IEEEauthorrefmark{4} \textit{Tsinghua University}
\\
\{2120240691@mail., 2110694@mail., sunyu@, zhhaiwei@, yuanxj@\}nankai.edu.cn, sxsong@tsinghua.edu.cn}
}

\maketitle

\begin{abstract}
Inconsistent values are commonly encountered in real-world applications, which can negatively impact data analysis and decision-making. While existing research primarily focuses on identifying the smallest removal set to resolve inconsistencies, recent studies have shown that multiple minimum removal sets may exist,
making it difficult to make further decisions. While some approaches use the \textcolor{black}{most frequent values} as the guidance for the subset repair, this strategy has been criticized for its potential to inaccurately identify errors. To address these issues,  we consider the dependencies between attribute values to determine a more appropriate subset repair. Our main contributions include (1) formalizing the optimal subset repair problem with attribute dependencies and analyzing its computational hardness; (2) computing the exact solution using integer linear programming; (3) developing an approximate algorithm with performance guarantees based on cliques and LP relaxation; and (4) designing a probabilistic approach with an approximation bound for efficiency. Experimental results on real-world datasets validate the effectiveness of our methods in both subset repair performance and  downstream applications.
\end{abstract}


\section{Introduction}

Inconsistent data present a common challenge in real-world applications, due to heterogeneous data integration \cite{DBLP:conf/icde/ChuIP13},
environmental interference \cite{DBLP:journals/pvldb/AbedjanCDFIOPST16}, and sensor malfunctions \cite{DBLP:journals/tkde/SunSY24}. The presence of such discrepancies can impact data analysis and the corresponding decision-making process \cite{DBLP:conf/icde/TongSDWC16,zhu2024relational}.


\textcolor{black}{To mitigate the issues caused by inconsistency, data repair methods are applied to restore the quality and reliability. These methods are critical as they directly influence the accuracy of downstream application and decision-making processes. Among the various data repair techniques, subset repair (S-repair) is one of the most foundational and widely-used approaches, which is aimed at finding a smallest removal set to resolve the inconsistency among a database.}
\textcolor{black}{In practice, S-repair plays a crucial role in a wide range of real-world tasks, including but not limited to data integration \cite{DBLP:conf/apweb/WangZZGX24}, deduplication \cite{DBLP:journals/ftdb/IlyasC15},  and consistent query answering (CQA) \cite{DBLP:journals/jcss/LivshitsKW21}. 
Furthermore, as highlighted in \cite{DBLP:journals/pvldb/MiaoCLGL20}, certain class of practical problems, such as designing procurement strategies and resource allocation, can be reduced to finding an optimal S-repair.
}

The majority of existing research focuses on identifying an S-repair with the fewest removals. \cite{DBLP:conf/pods/LivshitsKR18,DBLP:journals/pvldb/MiaoCLGL20,carmeli2024database}. However, \textcolor{black}{there may be multiple removal sets of equal minimum size (refer to Example \ref{eg:example-motivation2} for detailed statement), making it challenging to further select the most suitable one \cite{DBLP:journals/tkde/SunSY24}. Besides, this strategy also ignores the signals provided by the values \cite{DBLP:journals/pvldb/ProkoshynaSCMS15}. }
Consequently, frequent values are used to estimate correctness, \textcolor{black}{for instance, by assigning high confidence to tuples with high density \cite{DBLP:journals/tkde/SunSY24}, high pattern frequency \cite{DBLP:journals/pvldb/RezigOAEMS21}, and strong co-occurrence \cite{DBLP:journals/pvldb/RekatsinasCIR17}, or by assigning low confidence to outliers \cite{DBLP:conf/kdd/SongLZ15}.}
Regrettably, as \cite{DBLP:journals/jdiq/IlyasR22} points out, this strategy can result in an unfair cleaning result, 
where clean data may be mistakenly identified as erroneous due to their rarity and only frequent values are preserved. 
\textcolor{black}{
In relational database, relation among attribute values is more important. Focusing solely on value frequency or similarity is one-sided.}
\textcolor{black}{Some methods take distribution stability into account, aiming to keep the value distribution close to an ideal distribution \cite{DBLP:journals/pvldb/ProkoshynaSCMS15, DBLP:journals/pvldb/LiuSGGKR24}. While such strategy relies on an ideal distribution as input, which is relative to domain knowledge and not always obtainable.}
Machine learning based methods \cite{DBLP:journals/pvldb/RekatsinasCIR17, DBLP:conf/sigmod/MahdaviAFMOS019} leverage multiple existing signals simultaneously.
In fact, these approaches focus on integrating different signals into a unified framework, rather than designing new signals.

Notably, it is more reasonable to assess an attribute value's error by considering its dependencies with other attributes and tuples. When two tuples violate integrity constraints, the one deviating more from the attribute dependencies is more likely to be erroneous. Given multiple minimal removal sets\footnote{Please refer to the formal statement in Definition \ref{df:minimal removal set}.}, we select the optimal S-repair to maximize conformance to the attribute dependencies for the remaining tuples.

\begin{figure}[t]
  \centering 
  \vspace{-3pt}  
  \subfloat[2024 monthly household electricity bill in Shanghai]{\includegraphics[scale=0.23]{ 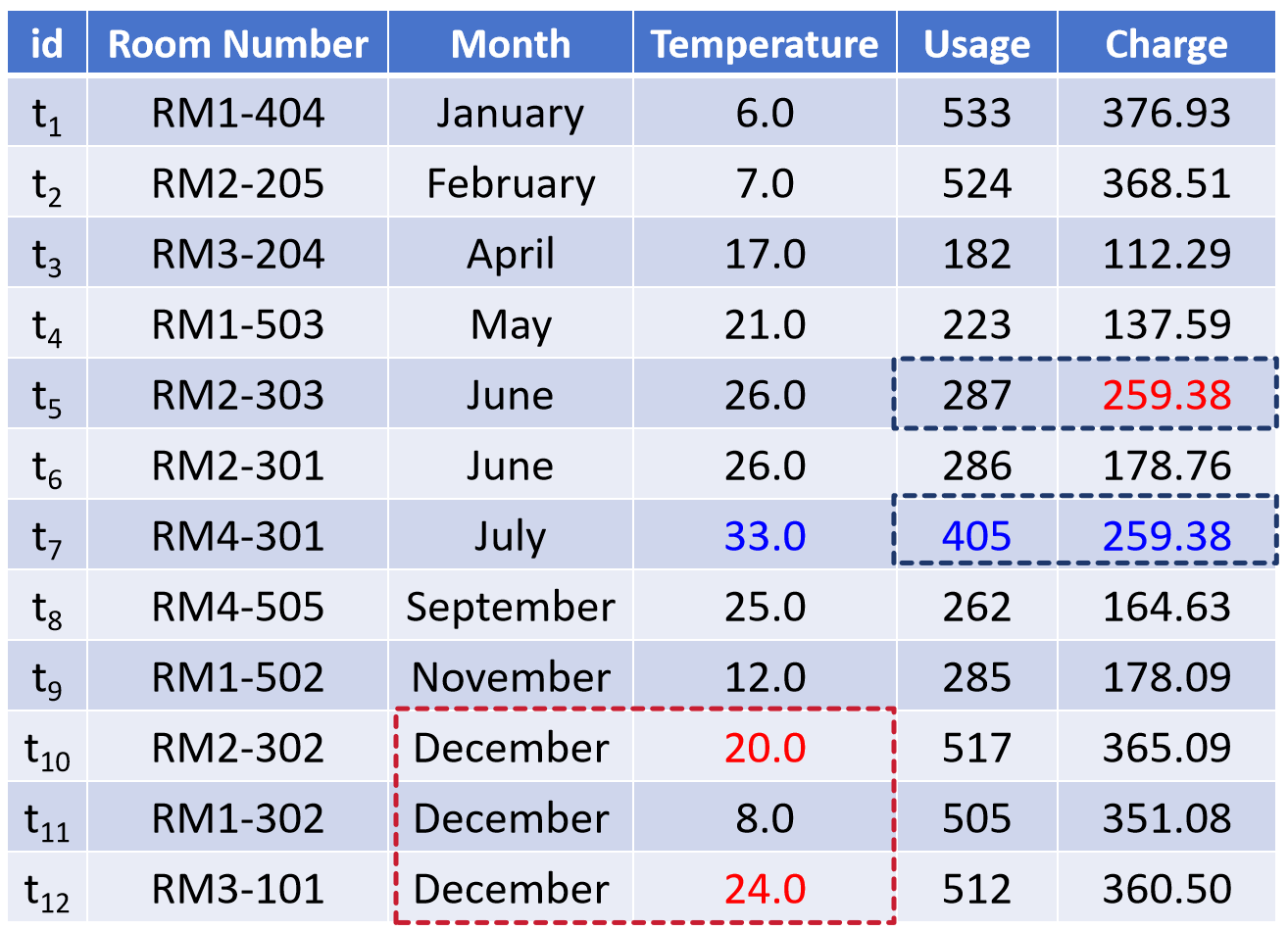} \label{fig:electricity_bill}} \\
  \vspace{-8pt}  
  \subfloat[\textcolor{black}{Average charges.}]{\includegraphics[scale=0.26]{ 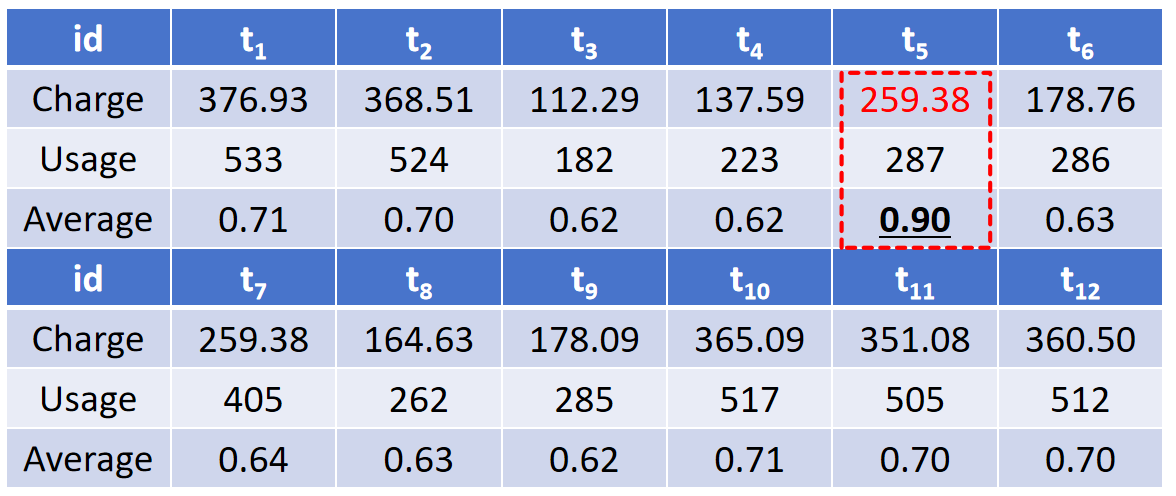} \label{fig:electricity_bill_avg_charge}} \\
  \vspace{-8pt}  
  \subfloat[\textcolor{black}{Tuple density calculated with the method of \cite{DBLP:journals/tkde/SunSY24}.} ]{\includegraphics[scale=0.32]{ 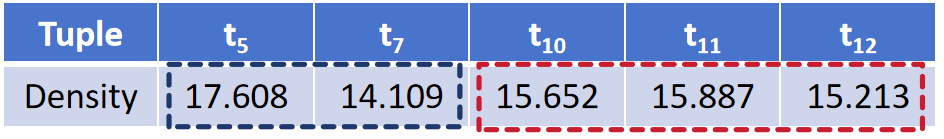} \label{fig:tp dst}} \\
  \caption{Example dataset. Values in \textcolor{red}{red} indicate errors, values in \textcolor{blue}{blue} are outliers.}
  \label{fig:combined-u-repair}
  \vspace{-10pt}  
\end{figure}

\begin{example}
\label{eg:example-motivation2}

\textcolor{black}{Figure \ref{fig:electricity_bill} presents the 2024 monthly household electricity bill in Shanghai, including Room Number, Month, Temperature (average temperature of the month), Usage (monthly electricity usage), and Charge.
Intuitively, tuples with the same Month should have the same temperature, and tuples with larger Usage should also have larger Charges. }

\textcolor{black}{In this instance, some conflicts exist: For tuples $t_{10}$, $t_{11}$, and $t_{12}$, they share the same Month value, but their Temperatures differ significantly. For $t_5$ and $t_7$, they have different Usages, but their Charges are identical. In Shanghai, the temperature in December averages around $5.7^\circ \text{C}$ \cite{ShangHai_temperature}, correlating with higher electricity consumption and charges. Therefore, $t_{10}$ and $t_{12}$, with Temperatures exceeding $20^\circ \text{C}$ in December, are likely errors. Regarding average electricity charges (as shown in \ref{fig:electricity_bill_avg_charge}), most values fall between 0.62 and 0.75, while $t_5$ has a value of 0.90, far outside this range, suggesting it is erroneous. Although errors can be detected this way, such strategies rely on domain knowledge, which may not always be available.}

\textcolor{black}{To solve such conflicts in an automated way, several strategies are optional. The first one is to find a minimum removal set to resolve the conflicts.} However there are six possible removal sets: $\{t_5,t_{10},t_{11}\}$, $\{t_5,t_{10},t_{12}\}$, $\{t_5,t_{11},t_{12}\}$, $\{t_7,t_{10},t_{11}\}$, $\{t_7,t_{10},t_{12}\}$, and $\{t_7,t_{11},t_{12}\}$. Determining the optimal solution among these six candidates is challenging without considering the attribute values. 
When using data frequency, \textcolor{black}{such as setting a high confidence for tuples that are close to others, (maximum density \cite{DBLP:journals/tkde/SunSY24}  with result shown in Figure \ref{fig:tp dst}}), the set $\{t_7,t_{10},t_{12}\}$ is selected as the output. Consider $t_5$ and $t_7$. $t_{7}$ is removed because it contains more outliers and a lower density. 
However, when considering dependencies within the values of $t_{7}$,  higher electricity usage correlates with a higher charge, which is more reasonable. Relying solely on a minimum repair or the most frequent values is inappropriate.  Therefore, we study attribute dependencies to better capture the relationships between attributes, aiming to identify a more reasonable removal set, $\{t_5,t_{10},t_{12}\}$.


\end{example}

\subsection{Challenges}
The problem of computing an optimal S-repair with the best conformance to the attribute dependency is challenging. 

(1) Since there may exist multiple tuple pairs violating the integrity constraints, to avoid excessive repair, there are a lot of candidate minimal removal sets, making the other remaining tuples compatible with each other. For example, in Figure \ref{fig:electricity_bill}, we need to consider all the conflicts between tuples and identify six minimal removal sets,  $\{t_5,t_{10},t_{11}\}$, $\{t_5,t_{10},t_{12}\}$, $\{t_5,t_{11},t_{12}\}$, $\{t_7,t_{10},t_{11}\}$, $\{t_7,t_{10},t_{12}\}$, $\{t_7,t_{11},t_{12}\}$. Therefore, it is not trivial to determine all the minimal removal sets w.r.t. the integrity constraints.

(2) The S-repair with the minimum removal tuples has already been proven to be NP-hard \cite{DBLP:journals/tkde/SunSY24}. Among multiple minimal S-repair solutions, it is thus even harder to compute the optimal S-repair conforming best to the attribute dependencies. That is, given six minimal removal sets in Figure \ref{fig:electricity_bill}, we need to further evaluate the conformance for each possible set w.r.t. attribute dependencies to determine the optimal one.

\subsection{Contributions}
Our major contributions in this study are as follows.

(1) We formalize the optimal S-repair problem w.r.t. attribute dependencies and integrity constraints, with the corresponding hardness analysis (Theorem \ref{th:hardness}) in Section \ref{sec:problem definition}.

(2) We compute the exact solution to the optimal S-repair problem in Section \ref{sec: exact method} with an integer linear programming.

(3) We devise an approximate method based on cliques and linear programming (LP) relaxation in Section \ref{sec:LP-relax} with approximate guarantee \textcolor{black}{(Proposition \ref{pp:cq method err bound})} and computational complexity  \textcolor{black}{(Proposition \ref{pp:cq method complexity})} analysis, whose optimality is ensured in certain cases.

(4) We design a probabilistic approach with approximation bound \textcolor{black}{(Proposition \ref{pp:prob method err bound})} in Section \ref{sec:prob based method} for high efficiency.

Extensive experiments over various real-world datasets demonstrate the superiority of our methods, on both S-repair performance and the downstream applications. 

\section{Problem Statement}
\label{sec:problem definition}
In this section, we introduce the subset repair under denial constraints in Section \ref{sect-subset-repair}. Then present the dependency models for attribute relations in Section \ref{sect-dependency-model}. The optimal subset repair problem is formally defined in Section \ref{sect-problem-definition}.

\subsection{Subset Repair}
\label{sect-subset-repair}

Consider a relation schema $R=\{A_1,A_2,...,A_m\}$, where each $A_j$, represents an attribute in $R$. A dirty instance $I$ of $R$ comprises a collection of tuples, denoted as $\{t_i\}_{i=1}^{n}$, each equipped with the attributes $\{A_1,A_2,...,A_m\}$. A tuple 
$t_i$ is written as $t_i=(t_i[A_1],t_i[A_2],$ $...,t_i[A_m])$. $t_i[A_j]$ signifies the value of tuple  $t_i$ on the attribute $A_j$. 

Integrity constraints are the most effective and widely adopted techniques to detect and repair errors \cite{DBLP:journals/pvldb/AbedjanCDFIOPST16,DBLP:conf/sigmod/SongZW16,DBLP:journals/pvldb/RekatsinasCIR17}, where the denial constraint (DC) is a popular one \cite{DBLP:conf/icde/ChuIP13}, owing to its expressive power. 
Denial constraints (DCs) are capable of capturing a broader array of relations, subsuming functional dependencies (FDs) \cite{DBLP:conf/kdd/MandrosBV17}, conditional functional dependencies (CFDs) \cite{DBLP:journals/tods/FanGJK08}, matching dependencies (MDs) \cite{DBLP:conf/icdt/BertossiKL11}, etc.
Therefore, in this study, we leverage DCs to detect the conflicts in the dirty instance $I$. A DC can be formulated as
\begin{equation}
    \varphi:\forall t_i, t_l \in I, \neg(P_{1} \wedge P_{2} \wedge ... ),
\end{equation}
where $P_{l}= (t_i[A] \phi t_l[A])$ is a predicate and $\phi\in \{>, \geq, <, \leq, =, \neq\}$ is an operator.    
A tuple pair $(t_i,t_l)$ is said to satisfy a predicate $P=(t_i[A_p] \phi t_l[A_q])$ if the relation $t_i[A_p] \phi t_l[A_q]$  holds true, denoted as $(t_i,t_l)\models P$.
The satisfaction of a constraint  $\varphi$ by the tuple pair $(t_i,t_l)$ is expressed as  $(t_i,t_l) \models \varphi$, indicating that there exists a predicate $P$ within $\varphi$ such that $(t_i,t_l)$ does not satisfy $P$.
For a given set $\Sigma$ of constraints, the tuple pair $(t_i,t_l)$ satisfies all the constraints $\varphi\in\Sigma$, we can say $(t_i,t_l)$ satisfy $\Sigma$, denoted as $(t_i,t_l) \models \Sigma$.
Furthermore, instance $I$ satisfies the set of constraints $\Sigma$, denoted as $I \models \Sigma$, if and only if all the tuple pairs  $t_i,t_l\in I$, satisfy $\Sigma$.

\textcolor{black}{\textbf{Note}: We assume the input constraints are suitable for repair, an assumption commonly used in rule-based repairing \cite{ DBLP:conf/icde/ChuIP13, DBLP:journals/tkde/SunSY24, DBLP:journals/pvldb/MiaoCLGL20}. For rule discovery, readers can refer to \cite{DBLP:journals/pvldb/XiaoTW022, DBLP:journals/pvldb/BleifussKN17}.}

To resolve conflicts within the dirty instance $\mathit{I}$, the subset repair (S-repair) typically identifies a subset of tuples to be removed, denoted as $I_N$, which guarantees that the remaining tuples $I\backslash I_N$ are free of violations. 
For a dirty instance $\mathit{I}$ over the schema $R$, with constraints $\Sigma$, $I_N\subset I$ is a removal subset if and only if $I\backslash I_N\models \Sigma$.
Designating all conflicting tuples as the removal set will undoubtedly eliminate all conflicts, yet it may cause an excessive repair. Therefore, we aim to find a minimal removal set of tuples to resolve conflicts.

\begin{definition}[Minimal Removal Set]
\label{df:minimal removal set}
    $I_N\subset I$ is a minimal removal set iff $\forall t_i\in I_N, \exists t_l \in I\backslash I_N, (t_i,t_l)\not \models \Sigma$. 
\end{definition}

Accordingly, the minimal condition excludes any tuple that, if reintroduced, would not cause a violation.
Moreover, the minimal removal set does not necessarily contain the minimum number of tuples to be removed (shown in Example \ref{eg:constraint and removal set}), and there may be several potential minimal removal sets available.

\begin{example} 
\label{eg:constraint and removal set}
Consider Figure \ref{fig:electricity_bill}. A DC is defined as:
\begin{equation}
\begin{split}
         \varphi_1:\forall t_i, t_l & \in I,   \neg(t_i[Month]=t_l[Month] \\ 
         & \wedge t_i[Temperature]\neq t_l[Temperature]),
\end{split}
\end{equation}
meaning that for any tuple pairs $t_i,t_l\in I$, if they satisfy 
$t_i[Month]$ $=t_l[Month]$ and $t_i[Temperature]\neq t_l[Temperature],$ they are involved in a conflict. Similarly, for the relations between $Usage$ and $Charge$, we can define:
\begin{equation}
    \begin{split}
         \varphi_2:\forall t_i, t_l \in I,  &  \neg(t_i[Usage]>t_l[Usage] \\ 
         & \wedge t_i[Charge]\leq t_l[Charge]).
    \end{split}
\end{equation}
The set of DCs in Figure \ref{fig:electricity_bill} is $\Sigma=\{\varphi_1, \varphi_2\}$. In Figure \ref{fig:electricity_bill}, $(t_{10},t_{11})$, $(t_{10},t_{12})$, $(t_{11},t_{12})$ violate $\varphi_1$ and $(t_{5},t_{7})$ violates $\varphi_2$. Among the tuples in conflict, $I_N=\{t_5,t_{10},t_{11}\}$ is a minimal removal set because putting back any $t_i\in I_N$ makes at least one conflict left.
\begin{figure}
    \centering
    \includegraphics[width=0.88\linewidth]{ 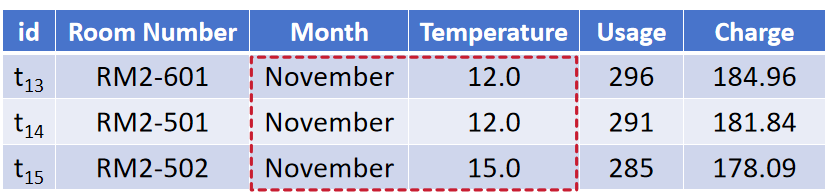}
    \caption{2024 monthly household electricity bill ($t_{13}-t_{15}$)}
    \label{fig:minimal and minimum}
    \vspace{-0.3in}
\end{figure}

\textcolor{black}{\textbf{Note:} A minimal removal set is distinct from a minimum removal set. As shown in Figure \ref{fig:minimal and minimum}, $(t_{13},t_{15})$ and $(t_{14},t_{15})$ violate $\varphi_1$. $\{t_{15}\}$ is a minimum removal set, as it has the smallest size. However, both $s_1=\{t_{13}, t_{14}\}$ and $s_2=\{t_{15}\}$ are minimal removal sets, because putting back any tuple from either $s_1$ or $s_2$ would leave a conflict in the remaining table.}
\end{example}
\vspace{-15pt} 

\subsection{Dependency Model}
\label{sect-dependency-model}
\vspace{-10pt} 
\textcolor{black}{Frequency value preferences, such as frequency, co-occurrence, and density, assume that a tuple $t_i$ is credible if there exists a group $G=\{t_l\}$ containing many similar or identical values with it. In other words, these principles require  $t_i$ to be close to $t_l \in G$. However, in a relational database, the similarity of attribute dependencies is more important than value similarity or distance. If the attribute dependencies in $t_i$ are similar to those in $t_l\in G$, regardless of their proximity, we consider $t_i$ to be reliable.
Therefore, we focus on attribute dependencies to evaluate tuple correctness.
}

Attribute dependencies are the relations between different attribute values. Given a tuple $t_i$  and an attribute $A_j$, we can estimate what $t_i[A_j]$ expected to be based on the other attribute values of $t_i$ with the dependencies. So we define dependency models to capture the relations.
Let $d_{ilj}$ denote the distance between $t_i[A_j]$ and $t_l[A_j]$. This distance can be computed using the absolute difference for numeric values and the Levenshtein distance for categorical values.
The dependency model $f_{lj}$ for $t_l\in I$ and $A_j\in R$,
\begin{equation}
\label{equation-dependency-model}
    d_{ilj}=f_{lj}(D_{ilj})+\varepsilon_{ilj}, 
\end{equation}
is used to predict $d_{ilj}$, where $\varepsilon_{ilj}$ is an error term and 
\begin{equation}
    D_{ilj}=[1,d_{il1},\cdots,d_{il,j-1},d_{il,j+1},...,d_{ilm}]
\end{equation}
is the distance vector for $t_i[A_j]$ and $t_l[A_j]$. 

\textcolor{black}{Dependency model $f_{lj}$ represents the relation of $t_l[A_j]$ with other attributes in $t_l$. If $t_i$ and $t_l$ have a similar relation between $A_j$ and other attributes, using $\hat{d}_{ilj}=f_{lj}(D_{ilj})$ to predict the distance of  $d_{ilj}$ should have a small residual. Actually, frequent values preference can be regarded as a special case of dependency model. When setting $d_{ilj}= 0 + \varepsilon_{ilj}$ in (\ref{equation-dependency-model}), the residual directly reflects the distance of $t_i[A_j]$ and $t_l[A_j]$.} 


As indicated in the existing study \cite{DBLP:conf/icde/ZhangSSW19}, for each tuple $\mathit{t}_{i}$, it is more accurate to consider the individual models between $\mathit{t}_{i}$ and its $\kappa$ nearest neighbors.
Therefore, for each dependency model $\mathit{f}_{lj}$, it captures the dependencies between tuple $\mathit{t}_{l}$ and its $\kappa$ nearest neighbors.
Consider the set of conflict tuples $I_C = \{t_i | \exists t_l \in I, (t_i, t_l) \not \models \Sigma\}$, which represents the tuples involved in conflicts. Let $N(t_l) \subset I \backslash I_C$ denote the set of $\kappa$ nearest neighbors of tuple $t_l$. Then the dependency model $\mathit{f}_{lj}$ could be learned between $t_l$ and $N(t_l)$.

\begin{example}
\label{eg:regression}
     \textcolor{black}{In Figure \ref{fig:electricity_bill}. 
     Suppose that the size of the training set $\kappa=4$. To obtain $f_{15}$, the $4-$nearest non-conflict tuples $t_2,t_4,t_6,t_9$ are leveraged. First, we calculate the distance vector between $t_i$ and $t_l$ ($i,l\in \{1,2,4,6,9\}$, $i<l$).}
     \begin{equation}
     \begin{split}
              D_{125}=[1.000, 0.352, &0.421, 0.037, 0.025]\text{,  }d_{125}=0.032;\\
              D_{145}=[1.000, 0.250, &0.696, 0.222, 0.706]\text{,  }d_{145}=0.751;\\
              &\cdots\\
              D_{695}=[1.000, 0.353, &0.600, 0.148, 0.179]\text{,  }d_{695}=0.156;\\
     \end{split}
     \end{equation}
     Fit $f_{15}$ with
     \begin{equation}
        f_{15}: 
        \begin{bmatrix}
        D_{125} \\ 
        D_{145} \\ 
        \cdots \\ 
        D_{695}
        \end{bmatrix}
        \to
        \begin{bmatrix}
        d_{125} \\ 
        d_{145} \\ 
        \cdots \\
        d_{695}
        \end{bmatrix}.
     \end{equation}
     For any tuple $t_i$ within the set $I$, $f_{15}$ can be utilized to predict the discrepancy between $t_1[Charge]$ and $t_i[Charge]$. This estimation is denoted as $\hat{d}_{i15}$. Suppose that $t_2$ is the given  $t_i$.  The distance vector:
     \begin{equation}
     \begin{split}
         D_{215}&=[1,d_{211},d_{212},d_{213},d_{214}]\\
         &=[1.000,0.352,0.421,0.037, 0.025].
     \end{split}
     \end{equation}
     Then the estimated distance $\hat{d}_{215}$ is calculated as 
     \begin{equation}
         \hat{d}_{215}=f_{15}(D_{215})=0.029.
     \end{equation}

\end{example}

    \textcolor{black}{\textbf{Note:} The regression model used in (\ref{equation-dependency-model}) can be any regression, such as polynomial regression \cite{DBLP:journals/iet-cds/HasanI19},  multi-layer perceptron \cite{popescu2009multilayer}, decision tree regression \cite{DBLP:journals/widm/Loh11}, etc. 
    In section \ref{sect-exp-proposed}, we estimate the sensitivity of methods to regression models. }

\subsection{Problem Definition and Analysis}
\label{sect-problem-definition}

In this section, we delineate the research problem addressed in our study. Considering tuples $t_i, t_l \in I$, according to (\ref{equation-dependency-model}), the dependency model $f_{lj}$ is employed to predict the distance $d_{ilj}$ between the attribute values $t_i[A_j]$ and $t_l[A_j]$.

The conformance loss $\ell(t_i,t_l;A_j)$ estimates the gap between the observed distance $d_{ilj}$ and the estimated $f_{lj}(D_{ilj})$, 
\begin{equation}
    \ell(t_i,t_l;A_j)=|d_{ilj}-f_{lj}(D_{ilj})|.
\end{equation}
 Considering all the dependency models on attributes $A_j\in R$, the overall conformance loss of $t_i$ is 
 \begin{equation}
     \ell(t_i,t_l)=\sum_{A_j\in R} \ell(t_i,t_l;A_j)
 \end{equation}

For each tuple $t_i \in I$, it is intuitive that only those sharing a similar attribute dependency model with $t_i$ yield a meaningful prediction loss $\ell(t_i,t_l)$. Tuples with different dependency relations would result in a significant prediction error. Consequently, we focus only on the subset of tuples that contribute the top-$k$ smallest $\ell(t_i,t_l)$ values for $t_i$, which we denote as $M(t_i)$. 


Considering all the tuples $\mathit{t}_{i}\in\mathit{I}$, the conformance loss is
\begin{equation}
    \ell(I)=\sum_{t_i\in I} \ell (t_i)=\sum_{t_i\in I}\sum_{t_l\in M(t_i)} \ell(t_i,t_l).
\end{equation}

However, focusing solely on minimizing $\ell(I\setminus I_N)$ may lead to excessive repair, because removing more tuples is a more efficient way to reduce $\ell(I\setminus I_N)$. To address this,  we normalize the prediction loss with a sufficiently large constant $G$,
\begin{equation}
\label{eq: Loss of the remaining set}
\begin{split}
    L(I\backslash I_N)&=\sum_{t_i\in I\setminus I_N} L(t_i)\\
    &=\sum_{t_i\in I\setminus I_N}\sum_{t_l\in M(t_i)}  L(t_i,t_l)\\
    &=\sum_{t_i\in I\setminus I_N}\sum_{t_l\in M(t_i)}  \sum_{A_j\in R}(G-\ell(t_i,t_l;A_j))\Gamma_i.
\end{split}
\end{equation}

The parameter $\Gamma_i$ serves to amplify the discrepancy between the tuple pair $(t_i, t_l)\not \models \Sigma$ that are in conflict with each other,
\begin{equation}
\Gamma_i=
    \begin{cases}
         \prod_{s=1}^{u_i} (1+\frac{\gamma}{s}) & \text{ if $u_i>0$,}\\
         1 & \text{ if $u_i=0$,}\\
        (\prod_{s=1}^{|u_i|} (1+\frac{\gamma}{s}))^{-1} & \text{ if $u_i<0$.}
    \end{cases}
\end{equation}
where $u_i=|\{t_l|\ell(t_i)<\ell(t_l), (t_i,t_l) \not\models \Sigma\}|-|\{t_l|\ell(t_i)>\ell(t_l),$$ (t_i,t_l)\not\models \Sigma\}|$ is to record how many times $t_i$ contains a larger or smaller $L(t_i)$ among the conflicts $t_i$ involved in. $\gamma>0$ is an enhancement constant. Maximizing $L(I\backslash I_N)$ balances model conformance and the number of repairs.

\begin{problem}
\label{df:problem}
Given an instance $I$ with the constraints $\Sigma$, the optimal S-repair (OSR) problem is to find a removal set $I_N$ of $I$, such that $I_N$ is minimal and maximizes $L(I\backslash I_N)$. 
\end{problem}


However, the OSR problem is difficult to solve.

\begin{theorem}
\label{th:hardness}
    The decision version of the OSR problem is NP-complete.
\end{theorem}
Proofs of all theorems and propositions are available at \cite{demo}.

\begin{example}
In Figure \ref{fig:electricity_bill}, let us consider a minimal removal set $I_N=\{t_5,t_{10},t_{12}\}$. With $k=2$ and $\gamma=0.2$, the conformance of the remaining dataset $I\backslash I_N$ can be computed using Equation (\ref{eq: Loss of the remaining set}), resulting in $L(I\backslash I_N)=10.023$. By evaluating the conformance for all possible removal sets $I_N$, we obtain the $L(I\backslash I_N)$ values for each corresponding remaining set. For instance, when $I_N=\{t_5, t_{10}, t_{11}\}$, the conformance is $L(I\backslash I_N)=9.458$, and when $I_N=\{t_7, t_{11}, t_{12}\}$, $L(I\backslash I_N)=8.532$. 
After calculating $L(I\backslash I_N)$ for all potential removal sets, the set $\{t_5,t_{10},t_{12}\}$ yields the highest $L(I\backslash I_N)$ value, indicating the best conformance to the attribute dependencies.
\end{example}

\section{Exact Algorithm}
\label{sec: exact method}
In this section, we transform the OSR problem into an Integral Linear Programming (ILP) formalization, to compute the exact solution.
For a given instance $I$, with a conflict tuple set $I_C=\{t_i|\exists t_j\in I, (t_i,t_j)\not\models \Sigma\}$, we introduce a decision variable $x_i$ for each tuple $t_i \in I_C$ . The variable $x_i=1$ if tuple $t_i$ is retained after repair, and $0$ otherwise. To encapsulate the condition of non-conflict, we ensure that at most one of the tuples $t_i$ and $t_l$ can be preserved if the pair $(t_i, t_l)$ does not satisfy $\Sigma$. Consequently, we impose the  constraint:
\begin{equation}
\label{cons:non-conflict}
    x_i+x_l\leq 1, \text{ (For $\forall (t_i,t_l)\not\models \Sigma$)}
\end{equation}

The weight $L(t_i)=\sum_{t_l\in M(t_i)} L(t_i,t_l)$ associated with a tuple $t_i$ is not static. It is influenced by the current set of top-$k$ providers and may change as certain tuples $t_l \in I_C$ are eliminated. To capture this dynamic relationship, we introduce a variable $y_{il}$ to indicate whether tuple $t_l$ continues to supply $L(t_i,t_l)$ to tuple $t_i$ post-repair. The variable $y_{il}=1$ if $t_l$ is among the top-$k$ providers for $t_i$, and $0$ otherwise. For tuple $t_i$ to be eligible to receive an  $L(t_i,t_l)$, it is necessary that both $t_i$ and $t_l$ are retained, which translates to $x_i=1$ and $x_l=1$. Therefore, we establish the following constraints:
\begin{equation}
\label{cons: xy rely 1}
\begin{split}
    &y_{il}\leq x_i, \text{ ($t_i\in I_C$, $t_l\in I$)}\\
    &y_{il}\leq x_l.  \text{ ($t_i\in I$, $t_l\in I_C$)}
\end{split}
\end{equation}

The constraint restricting  $t_i$ to receive no more than $k$ $L(t_i,t_l)$ is formulated as
\begin{equation}
\label{cons: topK1}
    \sum_{t_l\in \overline{M}(t_i)} y_{il} \leq k \text{ ($t_i\in I\backslash I_C$)}
\end{equation}
for non-conflict tuples and 
\begin{equation}
\label{cons: topK2}
    \sum_{t_l\in \overline{M}(t_i)} y_{il} \leq kx_i \text{ ($t_i\in I_C$)}
\end{equation}
for conflict tuples.  

$\overline{M}(t_i)$ is the set of potential $L(t_i,t_l)$ providers for $t_i$ during the repairing process, which can be calculated as follows.
Consider a $L(t_i,t_l)$ provider $t_l\in I\backslash I_C$ with 
\begin{equation}
    |\{t_s|L(t_i,t_l)\leq L(t_i,t_s), t_s\in I\backslash I_C\}|=k,
\end{equation}
its rank among the $L(t_i,t_l)$ providers is
\begin{equation}
\begin{split}
        K_i=|\{t_r| L(t_i,t_l)\leq L(t_i,t_r)\}|.
\end{split}
\end{equation}

Let $\overline{M}(t_i) = \{t_r \mid L(t_i,t_r) \geq L(t_i,t_l)\}$ be the set of tuples $t_r$ such that \(L(t_i,t_r) \geq L(t_i,t_l)\). The next proposition shows it is sufficient to only consider \(t_r \in \overline{M}(t_i)\) when setting \(y_{il}\).

\begin{proposition}
\label{$K_{i}$}
    For each $t_i\in I$, if $t_l\not \in \overline{M}(t_i)$, then $t_l$ can never provide top-$k$ $L(t_i,t_l)$ for $t_i$.
\end{proposition}

Therefore, we only generate \( y_{il} \) for \( t_l \in \overline{M}(t_i) \) of each \( t_i \). This reduces the number of \( y_{il} \) from \( O(n^2) \) to \( O(nK) \), where \( K = \max_i K_i \), which is typically much smaller than \( n \).

Combined with the objective of maximizing $L(I\backslash I_N)$, the ILP model for the OSR problem is 
\begin{equation}
\label{ILP OBJ2}
    \text{Max }\sum_{i=1}^n \sum_{t_l\in M(t_i)} y_{il}L(t_i,t_l)
\end{equation}
\begin{equation}
\nonumber
    \text{s.t. (\ref{cons:non-conflict}), (\ref{cons: xy rely 1}), (\ref{cons: topK1}), (\ref{cons: topK2})}
\end{equation}
$$x_i,y_{il}\in \{0,1\}$$

Suppose that $X^*$ is the solution returned by ILP. Then $X^*$ is the exact solution to OSR, as stated in the proposition below.

\begin{proposition}
     $X^*$ is the solution of the ILP. Then $I_N^*=\{t_i|x_i=0,x_i\in X^*\}$ is the exact solution to the OSR problem.
\end{proposition}

\begin{example}
\label{eg:ILP}
Consider the relation instance in Figure \ref{fig:electricity_bill}. First we find $\overline{M}(t_i)$ for each $t_i\in I$. For $t_1$, we could calculate $L(t_1,t_{10})=0.575$, $L(t_1,t_{11})=0.447$, $L(t_1,t_{12})=0.437$, $L(t_1,t_2)=0.375$, $L(t_1,t_8)$ $=0.027$, $\cdots$.
Among all potential providers, tuples $t_{10}, t_{11}, t_{12}$ belong to the conflict set $I_C$, while $t_2, t_8$ are from the non-conflict set $I\backslash I_C$. For any tuple $t_l$ where $L(t_1,t_l) < L(t_1,t_8)$, it is impossible for $t_l$ to serve as a provider for $t_1$. This is because $t_2, t_8$ would always precede $t_l$ as top-$k$ providers and would not be removed. Consequently, $\overline{M}(t_1)$ is set as $\{t_{10}, t_{11}, t_{12}, t_2, t_8\}$.

To formulate the Integer Linear Programming (ILP) model, we define variables $x_5, x_7, x_{10}, x_{11}, x_{12}$ to represent the removal of tuples $t_5, t_7, t_{10}, t_{11}, t_{12}$. The variables $y_{il}$ are used to indicate whether $t_l$ is an $L(t_i,t_l)$ provider after the repair, for all $t_i \in I$ and $t_l \in \overline{M}(t_i)$. To ensure a conflict-free condition, constraints such as $x_5 + x_7 \leq 1$ are imposed on all conflict pairs. Constraints like $y_{57} \leq x_{5}$ and $y_{57} \leq x_7$ are set to enforce the prerequisite for $y_{il} = 1$. For non-conflicting tuples, constraints like $\sum_{t_l \in \overline{M}(t_1)} y_{1l} \leq 2$ are used, and for conflicting tuples, constraints like $\sum_{t_l \in \overline{M}(t_5)} y_{5l} \leq 2x_5$ ensure that no tuple has more than $k$ providers. By incorporating these variables and constraints into the ILP, we obtain an exact solution with $x_5 = x_{10} = x_{12} = 0$ and $x_7 = x_{11} = 1$. Thus, the optimal removal set $I_N$ is identified as $\{t_5, t_{10}, t_{12}\}$.
\end{example}

\section{Approximate Solution via Cliques}
\label{sec:LP-relax}

Given the NP-completeness of the OSR problem, solving the ILP formalization to obtain the exact results is impractical for large-scale instances. Therefore, in this section, we relax the ILP problem into the LP form, which is polynomial-time solvable. In Section \ref{sect-LP-solutions}, we discover that the solution of the LP relaxation is mainly determined by variable $x_i$.  In Section \ref{sect-clique-constraint}, we find that a structure called clique is closely related to the fractional solution of LP, which is prevalent and capable of yielding a more stringent set of constraints. Through the incremental incorporation of clique-based constraints into the LP in Section \ref{sect-lp-clique}, we are able to obtain an approximate solution that comes with a performance guarantee.


\subsection{Solutions of LP Relaxation}
\label{sect-LP-solutions}

 

When working with an ILP model, a common approach is to relax the constraints on the variables from $x_i, y_{il} \in \{0,1\}$ to $x_i, y_{il} \in [0,1]$ and then round the resulting fractional solution to the nearest binary value, either 0 or 1. The LP formulation of the OSR problem possesses a unique structure that allows the constraints (\ref{cons:non-conflict}), (\ref{cons: xy rely 1}), (\ref{cons: topK1}), and (\ref{cons: topK2}) to be categorized into two distinct groups. Constraints (\ref{cons:non-conflict}) are responsible for ensuring non-conflict, while (\ref{cons: xy rely 1}), (\ref{cons: topK1}), and (\ref{cons: topK2}) are involved in calculating the sum of the top-$k$ $L(t_i,t_l)$ for $t_i \in I$.
We analyze the structure of the solution space to examine the interplay between the constraints and variables.

For a given LP formulation (\ref{ILP OBJ2}), $x_i, y_{il}$ represent the values of decision variables, and a solution $X=[x_i,y_{il}]$ is a vector consisting of $x_i, y_{il}$. The solution $X$ is considered feasible if it adheres to all constraints (\ref{cons:non-conflict}) to (\ref{cons: topK2}). The feasible region of the LP is the set of all feasible solutions $X$, and an extreme point is one of its vertices. When the value of $x_i$ is changed, the values of other variables $x_l, y_{il} \in X$ are correspondingly adjusted to yield a new optimal solution $X'=[x_i',y_{il}']$. The change in the objective value due to the modification of $x_i$ is calculated as $\sum_{t_i\in I} \sum_{t_l\in \overline{M}(t_i)}L(t_i,t_l)(y_{il}'-y_{il})$.


By analyzing the solution space of LP, we find a necessary condition of the solution $X$ returned by LP.

\begin{proposition}[Necessary Condition for LP Solution]
\label{pp:necessary cdt of LP}
Suppose that $X$ is a feasible solution. $X_P$ and $X_N$ are disjoint subsets of $X$.
For any element $x_l$ within $X$, the set $\{y_{l1}, y_{l2}, \ldots, y_{ls_l}\}$ encompasses all $y_{lr}$ values greater than zero, and $s_l$ denotes its size. Consider $\varepsilon > 0$, an arbitrarily small constant.  $X^+ = [x_i + \varepsilon, x_l - \varepsilon, x_r]$ and $X^- = [x_i - \varepsilon, x_l + \varepsilon, x_r]$ are two induced solutions, where $x_i \in X_P$, $x_l \in X_N$, and $x_r \in X \setminus (X_P \cup X_N)$. The solution $X$ can be turned into $X^+$ if $X^+$ is feasible, and similarly for $X^-$.

If $X$ is returned, for $\forall X_P,X_N\subset X$, there exists $x_l\in X$, either turning $X$ to $X^+$ or $X^-$  makes $kx_l-\sum_{r=1}^{s_l-1} y_{lr} > x_{s_l}$. 
\end{proposition}

Proposition \ref{pp:necessary cdt of LP} indicates that a necessary condition for a feasible solution of the LP is that altering the value of any $X_P,X_N$ results in some $x_l$ satisfying $kx_l - \sum_{r=1}^{s_l-1}y_{lr} > x_{s_l}$. As stated in \cite{vazirani2001approximation}, 
any solution returned by an LP model is an extreme point of its feasible region.
In our LP formulation, there are two types of extreme points.

\begin{definition}[$\mathcal{X}$-solution, $\mathcal{Y}$-solution]
\label{df:X,Y-solution}
    A feasible solution $X$ is an $\mathcal{X}$-solution iff $\{x_i\}\subset X$ forms an extreme point of the feasible region of (\ref{cons:non-conflict}). Otherwise, it is a $\mathcal{Y}$-solution. 
\end{definition}

Accordingly, an $\mathcal{X}$-solution relies  on $x_i\in X$ to form an extreme point of (\ref{cons:non-conflict}), and a $\mathcal{Y}$-solution is determined by $y_{lr}\in X$ to meet $kx_l-\sum_{r=1}^{s_l-1}y_{lr}=x_{ls_l}$.
It is evident that \textbf{the LP result is either an $\mathcal{X}$-solution or a $\mathcal{Y}$-solution.}

In practice, we find that the LP  could seldom return a $\mathcal{Y}$-solution, so we further analyze the precondition of $\mathcal{Y}$-solution.

\begin{proposition}[Necessary Condition for $\mathcal{Y}$-Solution]
\label{pp:necessary condition for Y-solution}
     Suppose that $X=[x_i,y_{lr}]$ is a $\mathcal{Y}$-solution. Then,
     
    (1) $\exists X_P,X_N\subset X$, such that both $X^+$ and $X^-$ (led by $X_P,X_N$ according to Proposition \ref{pp:necessary cdt of LP}) satisfy constraints (\ref{cons:non-conflict}). 

    (2) Consider $X^+$. Suppose that $x_i^+,y_{lr}^+\in X^+$ and 
    \begin{equation}
    \begin{split}
        T^+=\{t_l\mid & (kx_l^+ - \sum_{r=1}^{s_l-1}y_{lr}^+) - (kx_l - \sum_{r=1}^{s_l-1}y_{lr}) > 0, \\
        & kx_l - \sum_{r=1}^{s_l-1}y_{lr} = x_{s_l}\},\\
        T^-=\{t_l\mid & (kx_l^+ - \sum_{r=1}^{s_l-1}y_{lr}^+) - (kx_l - \sum_{r=1}^{s_l-1}y_{lr}) < 0, \\
        & kx_l - \sum_{r=1}^{s_l-1}y_{lr} = x_{s_l}\}.
    \end{split}
    \end{equation}
Let $\triangle y_{lr}=y_{lr}^+-y_{lr}$ be the value change of $y_{lr}$ caused by changing $X$ to $X^+$. Set 
\begin{equation}
\nonumber
\begin{split}
    \alpha&=\sum_{t_l\in I}\sum_{r=1}^{s_l-1} \triangle y_{lr} L(t_l,t_r)+\sum_{t_l\in I\backslash (T^-\cup T^+)}\triangle y_{lr}L(t_l,t_{s_l})\\
    &+\sum_{t_l\in (T^+\cup T^-)}(kx_l-\sum_{r=1}^{s_l-1}y_{lr})L(t_l,t_{s_l}),
\end{split}
\end{equation}
If $X$ is returned as a $\mathcal{Y}$-solution, for $X^+$ obtained from arbitrary $X_P$ and $X_N$, the following two formulas are satisfied.
\begin{equation}
\label{eq:complex cdt1-2}
    \alpha - \sum_{t_l\in T^+}(\triangle y_{ls_l}L(t_l,t_{s_l})-\triangle y_{l,{s_l}+1}L(t_l,t_{s_l+1}))\leq 0
\end{equation}
\begin{equation}
\label{eq:complex cdt2-2}
    -\alpha - \sum_{t_l\in T^-}(\triangle y_{ls_l}L(t_l,t_{s_l})-\triangle y_{l,{s_l}+1}L(t_l,t_{s_l+1}))\leq 0
\end{equation}
\end{proposition}
In Proposition \ref{pp:necessary condition for Y-solution}, condition (1) is prevalent because if no such $X_P,X_N$ exists, all $x_i\in X$ form an extreme point of the feasible region of (\ref{cons:non-conflict}). As for  condition (2),
if the solution $X$ does not adhere to Proposition \ref{pp:necessary cdt of LP}, the change in the objective value when transitioning from $X$ to $X^+$ is $\alpha$, and the change from $X$ to $X^-$ is $-\alpha$. However, if Proposition \ref{pp:necessary cdt of LP} holds, altering $X$ to $X^+$ would result in a set $T^+$ of tuples $t_l$ for which $kx_l - \sum_{r=1}^{s_l-1}y_{lr} > x_{s_l}$, thereby violating constraints (\ref{cons: xy rely 1}). The excess value $kx_l - \sum_{r=1}^{s_l-1}y_{lr} - x_{s_l}$ should be compensated by $y_{l,s_l+1}$. The value of $\alpha$ is calculated under the premise that no tuple $t_l$ satisfies $kx_l - \sum_{r=1}^{s_l-1}y_{lr} > x_{s_l}$, hence $\alpha$ exceeds the actual objective value change by $\sum_{t_l\in T^+}(\triangle y_{ls_l}L(t_l,t_{s_l}) - \triangle y_{l,{s_l}+1}L(t_l,t_{s_l+1}))$, which is accounted for in (\ref{eq:complex cdt1-2}). The analysis is analogous to the transition from $X$ to $X^-$.
Throughout the process of objective value modification, $\alpha$ constitutes the primary component, driven by $\sum_{t_l\in I}\sum_{r=1}^{s_l} \triangle y_{lr} L(t_l,t_r)$, while $\sum_{t_l\in T^+}(\triangle y_{ls_l}L(t_l,t_{s_l}) - \triangle y_{l,{s_l}+1}$ $L(t_l,t_{s_l+1}))$ represents an additional part resulting from $\sum_{t_l\in T^+}(kx_l $ $ - \sum_{r=1}^{s_l-1}y_{lr} - x_{ls_l})(L(t_l,t_{s_l}) - L(t_l,t_{s_l+1}))$.
Typically, $\alpha$ has a greater absolute value compared to the latter term. When (\ref{eq:complex cdt1-2}) yields a negative value, (\ref{eq:complex cdt2-2}) is usually positive, and vice versa. Consequently, the condition stipulated in Proposition \ref{pp:necessary condition for Y-solution} is uncommon, and the $\mathcal{X}-$ solution is predominantly returned.

\subsection{Clique Constraint}
\label{sect-clique-constraint}
As analyzed in Section \ref{sect-LP-solutions}, 
due to the rarity of Proposition 5, $\mathcal{X}$-solutions are returned most of the time, demonstrating the close relation between the $\mathcal{X}$-solution and the LP formula. So in this part, we study the properties of $\mathcal{X}$-solution.

$\mathcal{X}-$solution is mainly determined by the values of $x_i\in X$, which is closely related to the conflicts (\ref{cons:non-conflict}). 
Consider the feasible region $(P)$ led by the constraint in Formula (\ref{cons:non-conflict}).
\begin{equation}
\label{eq:feasible region (P)}
    \text{(P): }x_i+x_l\leq 1. \text{  $(t_i,t_l)\not\models\Sigma$}.
\end{equation}
The extreme points have the following characteristics.
\begin{proposition}
\label{pp:half integral}
     If $X$ is an $\mathcal{X-}$solution of LP, $\forall x_i, y_{il}\in X$, $x_i, y_{il}\in \{0,0.5,1\}^n$. 
\end{proposition}
The difference between the fractional solution and the integral solution is the variables $x_i\in X$ with $x_i=0.5$. If we can control the feasible region to eliminate the 0.5 values without influencing the values in $\{0,1\}$, an integral solution is derived. 
Due to the NP-completeness of the OSR problem, a polynomial time algorithm for general cases is unrealistic, but we can design a method with optimality in special cases.

Feasible region (P) is to achieve non-conflict, highly relying on the structure of conflicts.
In the field of constraint-based data repairing, conflicts detected by the integrity constraints are usually related to a structure called clique \cite{DBLP:journals/tkde/SunSY24,DBLP:journals/vldb/SongLCYC17}.
\begin{definition}[Clique]
    A conflict subset $I_q\subset I_C$ is a clique iff for any $t_i,t_l\in I_q$, $(t_i,t_l)\not \models \Sigma$.
\end{definition}






When there exists a clique $I_q$, $x_i=0.5$ is often observed. Consider a case that all the conflicts form a clique $I_q=\{t_1,t_2,...\}$, with fixed tuple weights $\{L(t_1), L(t_2),...\}$. Weight $L(t_i)$ is calculated by 
\begin{equation}
\label{eq:fixed weight}
    L(t_i)=\sum_{t_l\in M(t_i),M(t_i)\subset I}L(t_i,t_l).
\end{equation}
Then the following condition holds,

\begin{proposition}
    LP returns a solution  $x_i=0.5$ ($\forall t_i\in I_q$), if
    \begin{equation}
    \label{eq: clique 0.5 condition}
        \sum_{t_i\in I_q}x_iL(t_i)=0.5\sum_{t_i\in I_q} L(t_i)\geq max_{t_i\in I_q}L(t_i).        
    \end{equation}
\end{proposition}
This proposition indicates that 
for a clique $I_q$, if there is no $t_i\in I_q$ with an obviously greater weight than the others or the clique has a large size, LP will return a solution $X$ with $x_i=0.5$ (for $t_i\in I_q$).
In an integral solution, there is at most one $t_i\in I_q$ be remained. But the non-conflict constraints in Formula (\ref{cons:non-conflict}) cannot express such a relation, leading to $\sum_{t_i\in I_q}x_i^F>\sum_{t_i\in I_q}x_i^I=1$. So we introduce clique constraints, to tighten $x_i=0.5$ values.

\begin{definition}[Clique Constraint]
    For a clique $I_q$, the corresponding clique constraint is
    \begin{equation}
        \sum_{t_i\in I_q}x_i\leq 1.
    \end{equation}
\end{definition}
Compared to the constraints $x_i+x_l\leq 1$ $(t_i,t_l)\not\models \Sigma$, clique constraints are much tighter. It could force the sum of $x_i$ for $t_i\in I_q$ to be no more than 1.
As for an observed clique $I_q$, which has two subset $I_{q_1}$ and $I_{q_2}$, i.e., $I_q=I_{q_1}\cup I_{q_2}$, $I_{q_1}\neq I_{q_2}$. If we establish clique constraints $\sum_{t_i\in I_{q_1}}x_i\leq 1$ and $\sum_{t_l\in I_{q_2}}x_l \leq 1$ for $I_{q_1}$ and $I_{q_2}$, 
there may exist a feasible solution with $t_i\in I_{q_1}\backslash I_{q_2}$, $t_l\in I_{q_2}\backslash I_{q_1}$ and $t_v\in I_{q_1}\cap I_{q_2}$ forming a new clique $I_q'=\{t_i,t_l,t_v\}$ with $x_i=x_l=x_v=0.5$. We need to set one more clique constraint for $I_q'$. To avoid such case, we introduce maximal clique.
\begin{definition}[Maximal Clique]
    For a clique $I_q$, it is maximal iff $\forall t_i\in I_C\backslash \{I_q\}$, there exists a $t_l\in I_q$, $(t_i,t_l)\models \Sigma$.
    
\end{definition}

Moreover, clique constraints are naturally satisfied by ILP.

\begin{proposition}
    For any clique $I_q$, adding the clique constraint $\sum_{t_i\in I_q} x_i\leq 1$ into the ILP doesn't impact its solution space.
\end{proposition}

The LP with clique constraints is a relaxation of the original ILP. When the LP returns an integral solution, it is exact. The following proposition shows that in some cases, the LP with clique constraints can yield an exact solution. We first define the structure for this special case.

\begin{definition}[Chain]
    A chain is a sequence of tuples $I_s=(t_1,t_2,$ $...,t_s)$ satisfies $\forall t_i,t_j\in I_s$, $i\neq j$. 
\end{definition}

\begin{definition}[Disjoint Clique]
    $I_{q}$ is a disjoint clique if for all cliques $I_{q'}\subset I$, $I_q\cap I_{q'}=\emptyset$ or $I_{q'}=I_q$.
\end{definition}
The following proposition states the certain case optimality.

\begin{proposition}
\label{pp:optimality}
    Suppose the conflicts consist of chains and disjoint cliques. After adding clique constraints for $\forall I_q \subset I_C$ to the LP, if the LP solution $X$ is an $\mathcal{X}$-solution, it matches the original ILP solution exactly.

\end{proposition}

\begin{example}
Consider Figure \ref{fig:electricity_bill}, where the tuple set $I_q=\{t_{10},t_{11},$ $t_{12}\}$ constitutes a clique. Furthermore, it is a maximal clique because tuples $t_i \not\in I_q$ do not conflict with those in $I_q$.
Assume that the solution obtained from the LP relaxation is $x_{10}=x_{11}=x_{12}=0.5, x_5=0, x_7=1$. To tighten the constraints, a clique constraint $x_{10}+x_{11}+x_{12}\leq 1$ is incorporated into (\ref{cons:non-conflict}). Note that $t_5, t_7$ form a chain, and $I_q$ is a disjoint clique. According to Proposition \ref{pp:optimality}, if the LP yields an $\mathcal{X}$-solution $X$ after including the constraint $x_{10}+x_{11}+x_{12}\leq 1$, then $I_N=\{t_i | x_i=0, x_i \in X\}$ represents the exact solution to the ORS problem.
\end{example}



\begin{algorithm}
\caption{Clique($I$, $\Sigma$)}
\label{SolveWithClique}
\begin{algorithmic}[1]
    \Statex \textbf{Input:} a dirty instance $I$ with constraints $\Sigma$
    \Statex \textbf{Output:} a removal set $I_N$ of $I$
    \State $Q\gets \{\}$.\label{init}
    \While{ $max_{I_q\in Q_{new}}$ $|I_q|>2$ or $Q=\{\}$} \label{former loop states}
    \State $Q_{new}\gets \{\}.$ \label{init Qnew}
    \State Generate LP model with $I,L(t_i,t_l)$.  \label{LP building}
    \If {$Q\neq \emptyset$}
    \State $\forall I_q\in Q$, $|I_q|>2$,  add $\sum_{t_i\in I_q}x_i\leq 1$ to LP.\label{Clique conflict builder}
    \EndIf
    \State Solve LP. \label{LP solver}
    \State $I_h\gets \{t_i|x_i=0.5\}$\label{Ih building}
    \While{$\exists t_i\in I_h$, $t_i$ isn't marked}\label{CQ generating starts}
    \State $I_q\gets $ An maximum clique from unmarked $t_i\in I_h$.\label{build cliques}
    \State $Q_{new}.add(I_q)$.    \label{MC generating}
    \State Mark the $t_i\in I_q \cap I_h$.    \label{CQ generating ends}
    \EndWhile
    \State $Q\gets Q \cup Q_{new}$. \label{former loop ends}
    \EndWhile
    \State $I_N\gets \{t_i|x_i\leq 0.5\}$.\label{IN builder}
    \State Sort $t_i\in I_N$ in an descending order of $L(t_i)$.
    \For{$t_i\in I_N$}      \label{Check minimality starts}
    \If{$\forall t_l\in I\backslash I_N$, $(t_i,t_l)\models \Sigma$}
    \State $I_N.remove(t_i)$.   \label{Check minimality ends}
    \EndIf
    \EndFor
    \State \textbf{return } $I_N$.
\end{algorithmic}
\end{algorithm}

\subsection{Solving with Cliques}
\label{sect-lp-clique}


Based on the analysis in Sections \ref{sect-LP-solutions} and \ref{sect-clique-constraint}, the LP solution is determined by $x_i \in X$, with clique constraints converting fractional values to integers. In this section, we propose an approximation method with an error bound by incrementally adding clique constraints (Algorithm \ref{SolveWithClique}).


To start, $Q$ and $Q_{new}$ are initialized to store the cumulated and newly observed cliques.
The LP model is constructed in Line \ref{LP building} and subsequently solved in Line \ref{LP solver}. In Line \ref{Ih building}, variables with $x_i=0.5$ are aggregated into the set $I_h$. Cliques are then generated based on $I_h$, 
In Line \ref{MC generating}, an unmarked tuple $t_i$ with $x_i \in I_h$ is selected to initiate a clique $I_q$, with additional tuples $t_l \in I_C$ being incrementally added to $I_q$. Specifically, $t_l$ is included if and only if $(t_l, t_s) \not \models \Sigma$ for all $t_s \in I_q$, until no further qualifying $t_l$ outside $I_q$ can be found. During the subsequent iteration, in addition to using conflict pairs $(t_i, t_l) \not \models \Sigma$ to generate constraints, each clique $I_q \in Q$ will add a constraint of the form $\sum_{t_i \in I_q} x_i \leq 1$ (Line \ref{Clique conflict builder}). The loop persists until all newly observed cliques $I_q$ have a size $|I_q| \leq 2$.
Tuples $t_i$ with $x_i \leq 0.5$ are gathered into a removal set $I_N$ in Line \ref{IN builder}. From Lines \ref{Check minimality starts} to \ref{Check minimality ends}, redundant tuples are examined to ensure the minimality of $I_N$.

As shown in Algorithm \ref{SolveWithClique}, the method has an iterative framework. So we  analyze the convergency of Algorithm \ref{SolveWithClique}. 





\begin{proposition}
    Suppose that $c=|I_C|$ is the number of conflict tuples, Algorithm \ref{SolveWithClique} terminates within $C_c^3$ iterations.
\end{proposition}

Although the worst-case iteration $O(c^3)$ is high, the convergence speed in practice is usually very fast. We verify it in Section \ref{sect-exp-proposed} and report the number of iterations of Algorithm \ref{SolveWithClique} on all the datasets in the experiments in Table \ref{tab:Number of Iterations of Clique on each Dataset}.

Then, we verify that Algorithm \ref{SolveWithClique} always returns a solution meeting the minimality condition in polynomial time.

\begin{proposition}
\label{pp:cq method complexity}
    Algorithm \ref{SolveWithClique} returns a minimal removal set in $O(mn^2+(nK+c)^{3.5}c^3)$ time, where $\mathit{m}=|R|$, $\mathit{n}=|I|$, $\mathit{c}=|I_C|$, and $K=max_{t_i\in I}|\overline{M}(t_i)|$. 
\end{proposition}

Next, we study the approximation of Algorithm \ref{SolveWithClique}.

\begin{proposition}
\label{pp:cq method err bound}
    Suppose that $I_N^*$ is the exact solution. The $I_N$ found by Algorithm.$\ref{SolveWithClique}$ satisfies, 
    \begin{equation}
    \label{error bound}
        \frac{L(I\backslash I_N)}{L(I\backslash I_N^*)}\geq \eta \frac{|I\backslash I_C|}{n},
    \end{equation}
    where $\eta=\frac{minL(t_i,t_l)}{maxL(t_i,t_l)}$, $n=|\mathit{I}|$.
\end{proposition}

\begin{example}
\label{eg: solve with clique}

Consider Figure \ref{fig:electricity_bill}. Algorithm \ref{SolveWithClique} addresses the LP relaxation of the ILP from Example \ref{eg:ILP} during the first iteration, from Line \ref{init Qnew} to Line \ref{LP solver}. Suppose the LP solution includes $x_{10} = x_{11} = x_{12} = 0.5$, $x_5 = 0$, and $x_7 = 1$. We construct $I_h = \{x_{10}, x_{11}, x_{12}\}$ for the variables with value 0.5 (Line \ref{Ih building}). Between Line \ref{CQ generating starts} and Line \ref{CQ generating ends}, cliques $I_q=\{t_{10},t_{11},t_{12}\}$ are generated and added to $Q_{new}$ (Line \ref{MC generating}). Initially, all tuples in $I_h$ are unmarked. Starting with $t_{10}$, we initialize a clique $I_q = \{t_{10}\}$, and incorporate unmarked tuples $t_{11}$ and $t_{12}$, forming $I_q = \{t_{10}, t_{11}, t_{12}\}$, which is then added to $Q_{new}$. These tuples are marked in Line \ref{CQ generating ends}. In the next iteration, a new clique constraint $x_{10} + x_{11} + x_{12} \leq 1$ is added to (\ref{cons:non-conflict}) (Line \ref{Clique conflict builder}), and the revised LP is solved (Line \ref{LP solver}), yielding $x_7 = x_{11} = 1$, and $x_5 = x_{10} = x_{12} = 0$. The loop ends when no variable has a value of 0.5 (Line \ref{former loop states} to Line \ref{former loop ends}). The removal set $I_N = \{t_5, t_{10}, t_{12}\}$ is formed (Line \ref{IN builder}), and redundancy is checked for minimality (Line \ref{Check minimality starts} to Line \ref{Check minimality ends}).

\end{example}

\section{Probabilistic Approximation}
\label{sec:prob based method}
Although the LP-based method can find an approximate solution in polynomial time, the computational cost of $O(n^{3.5})$ is somewhat prohibitive. Consequently, a more efficient heuristic approach is necessary without calling the LP solver. 
Considering that a tuple $t_i$ with a higher $L(t_i)$ value is more likely to be accurate we utilize loss-based probability to resolve conflicts.

Suppose that $L(t_i)$ for $t_i\in I$ is pre-calculated and considered fixed.
 For each pair $(t_i,t_l)\not \models \Sigma$, either $t_i$ or $t_l$ is removed to eliminate the conflict. So the sum of their removal probability should be 1. Tuple with the higher $L(t)$ (suppose $L(t_i) > L(t_l)$) is more probable to be correct. Therefore,  we can assign a higher probability of retaining this tuple:
 

\begin{equation}
\label{eq:Pil}
    P_{il}=\frac{L(t_i)}{L(t_i)+L(t_l)},
\end{equation}
$P_{il}$ is the probability of keeping $t_i$. In the opposite, the probability of keeping $t_l$ is,
\begin{equation}
\label{eq:Pli}
    P_{li}=\frac{L(t_l)}{L(t_i)+L(t_l)}.
\end{equation}
Therefore, we can design a probabilistic algorithm to find $I_N$:

\begin{algorithm}
\caption{Probabilistic($I$, $\Sigma$)}
\label{Random}
\begin{algorithmic}[1]
\Statex \textbf{Input:} a dirty instance $I$ with constraints $\Sigma$
\Statex \textbf{Output:} a removal set $I_N$ of $I$
\State $I_N\gets \{\}$. \label{IN builder 2}
\State Set $P_{il}$ and $P_{li}$ for $\forall (t_i,t_l)\not\models \Sigma$ based on (\ref{eq:Pil}) (\ref{eq:Pli}).\label{Set Probabilistic}
\State $I_N.add(t_i)$ with probability $1-P_{il}$ or otherwise $I_N.add(t_l)$.\label{Probabilistic ends}
\State Sort $t_i\in I_N$ in an descending order of $L(t_i)$. \label{sort IN 2}
\For{$t_i\in I_N$} \label{Check Minimal starts 2}
\If{$\forall t_l\in I\backslash I_N$, $(t_i,t_l)\models \Sigma$}\label{check minimal2}
\State $I_N.remove(t_i)$. \label{Check Minimal ends 2}
\EndIf
\EndFor
\State \textbf{return} $I_N$.
\end{algorithmic}
\end{algorithm}

Probabilistic first assigns retaining  probabilities $P_{il}$ and $P_{li}$ to each pair $(t_i, t_l) \in I_C$  in Line \ref{Set Probabilistic}. 
The conflict pairs are considered independently. $t_i$ may be added into set $I_N$ repeatedly when considering different conflicts. 
In Line \ref{Probabilistic ends}, it incorporates either $t_i$ or $t_l$ into $I_N$ with the respective probabilities of $1-P_{il}$ and $1-P_{li}$.
From Lines \ref{Check Minimal starts 2} to \ref{Check Minimal ends 2}, the redundancy of each tuple within $I_N$ is assessed to ensure the minimal condition. Because the $t_i$ with a larger $L(t_i)$ is more likely to be correct, in Line \ref{sort IN 2} we sort the $t_i\in I_N$ in descending order to prioritize tuples with larger $L(t_i)$. In Line \ref{Check Minimal starts 2} the $t_i\in I_N$ are traversed and its redundancy is checked in Line \ref{check minimal2}. The redundant tuples are put back in Line \ref{Check Minimal ends 2}.

As is shown, Algorithm \ref{Random} has a concise process, which makes it more efficient than Algorithm \ref{SolveWithClique}.
\begin{proposition}
\label{pp:prob method complexity}
    Algorithm \ref{Random} returns a minimal removal set  $I_N$ in $O(mn^2)$ time. Where $n=|I|$, $m=|R|$.
\end{proposition}

Although Algorithm \ref{Random} has a simple framework, it owns the valuable approximation performance guarantee.

\begin{proposition}
\label{pp:prob method err bound}
    Suppose that $I_N^*$ is the exact solution. $I_N$ found by Algorithm.$\ref{Random}$ satisfies, 
    \begin{equation}
        \frac{E(L(I\backslash I_N))}{E(L(I\backslash I_N^*))}\geq (\frac{\eta}{2})^{2V+1}.
    \end{equation}
    $V$ is the maximum conflicts a tuple $t_i\in I_C$ is involved in. $\eta=\frac{minL(t_i,t_l)}{maxL(t_i,t_l)}$. And $E(L(I\backslash I_N))$ and $E(L(I\backslash I_N^*))$ denotes the expectation of $I\backslash I_N$ and $I\backslash I_N^*$.
\end{proposition}


\begin{example}
Consider the example depicted in Figure \ref{fig:electricity_bill}. Assume that the loss values are calculated as $L(t_5)=0.005$, $L(t_7)=1.220$, $L(t_{10})=0.813$, $L(t_{11})=1.213$, and $L(t_{12})=0.525$ with (\ref{eq: Loss of the remaining set}). The conflict pairs are examined to assign retention probabilities. For the pair $(t_5,t_7)$, the retention probabilities are computed as $P_{57}=\frac{0.005}{0.005+1.220}=0.004$ and $P_{75}=\frac{1.220}{0.005+1.220}=0.996$. Consequently, with a probability of $1-P_{57}=0.996$, $t_5$ is removed, or $t_7$ is removed otherwise.
This process is repeated for all conflict pairs. Suppose that the resulting removal set after this process is $I_N'=\{t_5,t_{10},t_{11},t_{12}\}$, it will be sorted as $[t_{11},t_{10},t_{12},t_5]$ in Line \ref{sort IN 2} and check minimality in Lines \ref{Check Minimal starts 2} to \ref{Check Minimal ends 2}. Finally, $t_{11}$ is reinserted, and  $I_N=\{t_5,t_{10},t_{12}\}$ is returned as the outcome.
\end{example}

\section{Experiments}
\label{sec:exp}
In this section, we carry out an experimental analysis to ascertain the following: (1) The performance of our methods in comparison to eight competing methods, across both real and synthetic error scenarios.   
(2) The practical convergence speed of the clique-based method and 
the parameter sensitivity of our algorithms, 
including the enhancement parameter $\gamma$, training set size $\kappa$, number $k$ of $L(t_i,t_l)$ for calculating $L(t_i)$, number $m$ of models selected for training $f_{ij}$ ($t_i\in I,$ $A_j\in R$), \textcolor{black}{type of the regression models}.  
(3) Whether our methods can confer superior benefits to downstream classification applications in comparison to other data-cleaning methodologies.

The source code and data are available online \cite{demo}. 

\subsection{Experimental Setting}
Before presenting the experiment results, we introduce the datasets, metrics, and baselines used in our empirical study.
\subsubsection{Datasets}
In experiments, we utilize two datasets Flights, Rayyan with real-world errors and 
Soccer, Restaurant, Income and Inspection with synthetic errors to evaluate the S-repair performance, as well as Iris and Yeast with embedded errors by KEEL \cite{keel} for the application study.



\textbf{Flights} \cite{DBLP:journals/pvldb/LiDLMS12} includes records of scheduled and actual arrival and departure times for various flights. \textcolor{black}{It is a composite dataset from multiple sources, resulting in some conflicts for the same flight in both scheduled and actual times.} These errors have been identified by the dataset providers.

\textbf{Rayyan} \cite{ouzzani2016rayyan} includes metadata about article titles, authors, languages, journals, etc. Some details of attribute values in this dataset are represented by the character ('?'), and there are instances where tuples are incomplete due to missing values. These errors stem from formatting issues and have been rectified by the dataset developers.



\textbf{Soccer}, \textbf{Restaurant}, \textbf{\textcolor{black}{Income}} and \textbf{Inspection} \cite{DBLP:journals/pvldb/ArocenaGMMPS15,DBLP:journals/pvldb/XiaoTW022} are widely used datasets in existing data-cleaning studies \cite{DBLP:journals/tkde/SunSY24,DBLP:journals/pvldb/ArocenaGMMPS15,DBLP:journals/pvldb/LiuSGGKR24}, which are considered originally clean. We employ FastADC \cite{DBLP:journals/pvldb/XiaoTW022} to identify the DCs and utilize \textcolor{black}{a benchmark error injection tool} Bart \cite{DBLP:journals/pvldb/ArocenaGMMPS15} to generate synthetic errors, including \textcolor{black}{typos, duplicated values, bogus values, outliers}. The information related to these datasets is shown in Table \ref{tab:dataset information}.


\begin{table}[t]
\caption{Dataset characteristics}
\vspace{-0.1in}
\resizebox{8cm}{!}{
\label{tab:dataset information}
\vspace{-0.15in}
\centering
\begin{tabular}{l|cccccc}
\hline
Dataset & Size & Attribute & DCs & FDs & Duplicate & Class \\ \hline
Rayyan & 1000 & 10 & 4  & 1  & Few & - \\ 
Flights & 2376 & 6 & 4  & 3  & Few & - \\ 
Restaurant & 800 & 4 & 2  & 1  & Few & - \\ 
Soccer & 1600 & 7 &  3  & 3  & Many & - \\ 
Inspection & 200k & 13 & 15 & 6  & Medium & - \\ 
\textcolor{black}{Income} & 1M & 13 & 4 & 2 & Medium & - \\
Iris & 150 & 4 & 2 & 1  & Few & 3 \\
Yeast & 1484 & 8 & 4  & 3  & Few & 10 \\ \hline
\end{tabular}
}
\vspace{-0.2in}
\end{table}



\textbf{Iris} and \textbf{Yeast} \cite{keel} are employed for classification tasks. 
Both datasets are infused with errors ranging from 5\% to 20\%. These errors have been introduced by the dataset providers.

\subsubsection{Metrics}
To compare the accuracy of different methods, precision, recall and $F1-score$ are used. Precision measures the accuracy of removal, calculated as $precision=\frac{|Truth\cap Removal|}{|Removal|}$, where $Truth$ is the set of tuples with true errors and $Removal$ is the returned removal set. Recall is calculated as $recall=\frac{|True\cap Removal|}{|True|}$, representing the algorithm's coverage of errors. $F1-score=2\frac{precision\cdot recall}{precision+recall}$ is the combination of the above metrics.

\subsubsection{Baselines}

We compare our methods \textbf{Clique} and \textbf{Probabilistic}, \textcolor{black}{setting linear regression for dependency models,} with eight widely used data-repairing approaches.
\textcolor{black}{\textbf{TE-LP} \cite{DBLP:journals/pvldb/MiaoCLGL20} is a rule-based S-repair method that focuses on the minimum principle.
\textbf{OptSRepair} \cite{DBLP:journals/tods/LivshitsKR20} is a cleaning method leveraging value frequency and maximum matching for repairing.
\textbf{Soft} \cite{carmeli2024database} is a method considering the correctness of constraint, allowing repaired data to violate rules at a certain cost.}
\textbf{Relaxation} \cite{DBLP:journals/tkde/SunSY24} is a density-based S-repair method, which sets density as the signal for error detection. 
\textbf{Holistic} \cite{DBLP:conf/icde/ChuIP13} is a rule-based method compiling  DC conflicts into a hypergraph and applies minimum vertex cover  to find the erroneous cells. 
\textbf{Holoclean (HC)} \cite{DBLP:journals/pvldb/RekatsinasCIR17} is a machine learning method combining both integrity constraints and probabilistic inference for error detection and repair.  
\textbf{Raha} \cite{DBLP:conf/sigmod/MahdaviAFMOS019} is another machine learning system integrating multiple data-cleaning algorithms.
\textbf{Horizon} \cite{DBLP:journals/pvldb/RezigOAEMS21} is a near-linear time method based on the most frequent patterns. 
Considering that DCs are applied for cleaning, \textcolor{black}{we only set the FDs for methods solving with FD conflicts, including Soft, OptSRepair, Horizon.} For some of the update-repair (U-repair) methods, i.e., Holistic, Holoclean and Horizon, we select the tuples with cells to modify as removal set.

\textcolor{black}{\textbf{Note: }\textbf{ILP} is the method to calculate the exact solution of problem \ref{df:problem}, which is NP-complete. So this method is not scalable, which is the reason of designing Clique and Probabilistic. ILP is a theoretical method to study the correctness and sensitivity of problem definition in section \ref{sect-exp-proposed}. }

\subsection{Comparison on Real Errors}
\label{sec:exp-real-error}

\begin{table}[t]
\centering
\setlength{\tabcolsep}{1.5pt}
\caption{Performance on Flights and Rayyan datasets}
\resizebox{8cm}{!}{
\begin{tabular}{c|cccc|cccc}
\hline
& \multicolumn{4}{c|}{Flights} & \multicolumn{4}{c}{Rayyan} \\
\hline
Methods & Pre & Rec & F1 & Time (s) & Pre & Rec & F1 & Time (s) \\
\hline
\textcolor{black}{TE-LP} & 0.788 & 0.633 & 0.702 & 0.791 & 0.819 & 0.766 & 0.792 & 0.996 \\
\textcolor{black}{OptSRepair} & 0.934 & 0.515 & 0.664 & 0.059 & 0.964 & 0.428  & 0.593 & 0.047 \\
\textcolor{black}{Soft} & 0.706 & 0.441 & 0.543 & 8.611 & 0.937 & 0.447 & 0.605 & 5.200 \\
Relaxation & 0.86 & 0.744 & 0.798  & 0.481 &  0.717 & 0.902 & 0.799 & 0.465 \\
Holistic & 0.802 & 0.664 & 0.726 & 0.389 & 0.709 & 0.893 & 0.791 & 0.348 \\
Horizon & 0.763 & 0.605 & 0.675 & 0.045 & 0.772 & 0.445 & 0.564 & 0.043 \\
Raha & 0.642 & 0.986 & 0.777 & 2.158 & 0.804 & 0.832 & 0.817 & 1.038 \\
HC & 1.000 & 0.070 & 0.131 & 21.147 & n/a & n/a & n/a & 48.032 \\
Probabilistic & 0.898 & 0.845 & 0.870 & 0.128 & 0.747 & 0.940 & 0.832 & 0.093 \\
Clique & 0.905 & 0.847 & \textbf{0.875} & 32.025 & 0.749 & 0.943 & \textbf{0.834} & 14.931 \\
\hline
\end{tabular}%
}
\label{tab:results on real-world errors}%
\vspace{-0.2in}
\end{table}%

We first compare our algorithms against baselines over the real-world erroneous datasets Flights and Rayyan in Table \ref{tab:results on real-world errors}.
Among the constraint-based methods, \textcolor{black}{OptSRepair, Soft and} Horizon demonstrate the lowest accuracy, due to the drawbacks of FDs. This same limitation is responsible for its suboptimal performance as depicted in Figures \ref{fig:restaurant}, \ref{fig:income}, \ref{fig:inspection}.   
However, with a linear complexity, Horizon achieves the quickest cleaning time. On both datasets, Relaxation outperforms TE-LP and Holistic, an advantage stemming from its ability to make further decisions based on density. 
Raha's performance is unstable, as its accuracy depends heavily on the quality of sampling and labeling. As for HC, the original values are used as labels for training unless there is sufficiently high confidence to set new values. This results in only a small subset of cells being repaired with new values.
In the following experiments (in Figure \ref{fig:restaurant},  \ref{fig:income}, \ref{fig:inspection}), the result of HC is omitted as it fails to repair.
Among all the methods, our Clique stands out for its accuracy, a strength derived from the attribute dependencies. Probabilistic yields the second most accurate results after Clique, while also offering greater efficiency by its concise framework.

\subsection{Comparison on Synthetic Errors}
\label{sec:exp-syn-error}

In this part, we compare the methods on datasets Soccer, Restaurant,  Income and Inspection with synthetic errors. For Soccer and Restaurant, we set error tuples ranging from 10\% to 40\%. Income with 1M tuples and Inspection with 200k tuples are used to test the scalability, so we uniformly add 15\% error tuples with Bart for both datasets and conduct experiments with $10\%$ to $100\%$ tuples.
 \begin{figure}[t]
  \centering
  \includegraphics[width=\linewidth]{ 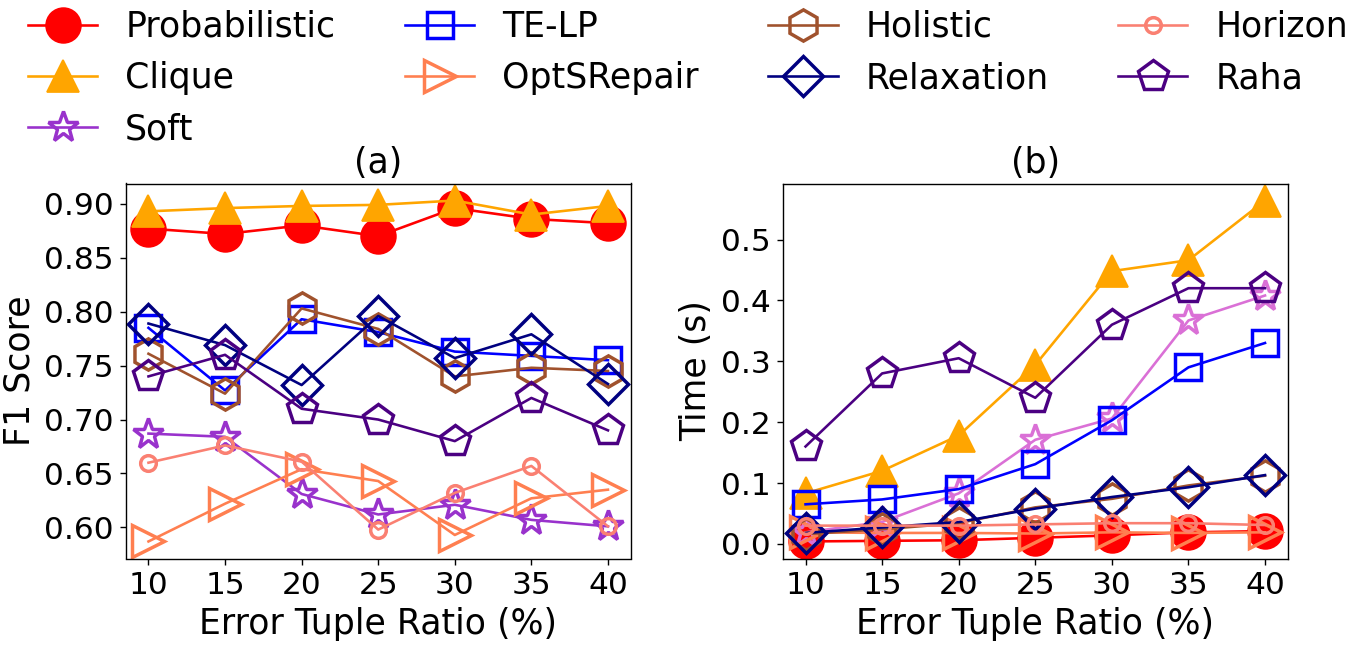} 
  \vspace{-0.25in}
  \caption{\textcolor{black}{Performance on Restaurant with varying error rates}}
  \label{fig:restaurant}
  \vspace{-0.15in}
\end{figure}

\begin{figure}[t]
  \centering
  \includegraphics[width=\linewidth]{ 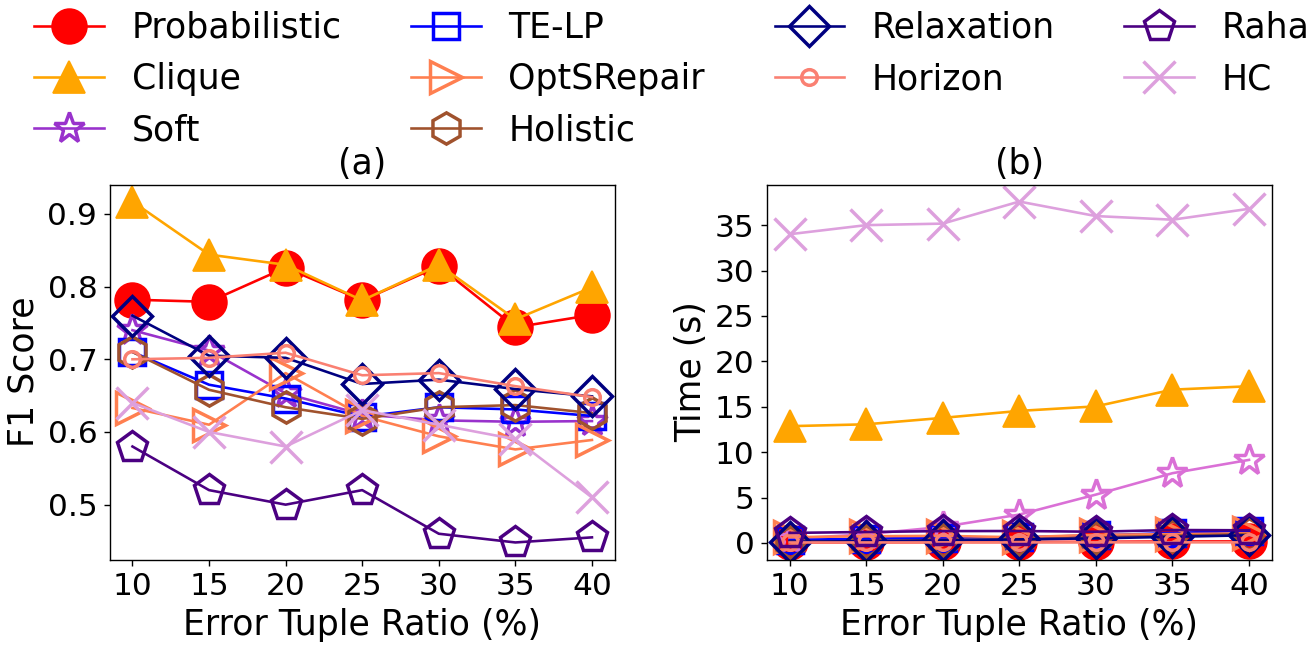} 
  \vspace{-0.25in}
  \caption{\textcolor{black}{Performance on Soccer with varying error rates}}
  \label{fig:soccer}
  \vspace{-0.15in}
\end{figure}

Figures \ref{fig:restaurant} and \ref{fig:soccer} illustrate the outcomes of various methods applied to the Restaurant and Soccer datasets. It is evident that as the error rate escalates, the difficulty of the repair tasks intensifies, leading to a decline in the accuracy of all methods.
As depicted in Figure \ref{fig:soccer}, Horizon exhibits performance comparable to Relaxation, primarily because the constraints employed in the Soccer dataset are all FDs, and most frequent patterns are somewhat similar to the highest density, i.e., they confer greater credibility to frequent values. 
Clique emerges as the most accurate among all the methods. The accuracy of Probabilistic is marginally lower than Clique, as Clique addresses conflicts from a more global perspective.

In terms of speed, a higher error ratio leads to more  conflicts or patterns for  resolution, prompting most methods to dedicate additional time to error correction. Owing to the high time complexity, the Clique method operates more slowly compared to other constraint-based techniques. 
In contrast, benefiting from the efficient framework, Probabilistic  outperforms most other methods in terms of speed, even rivaling  Horizon.

\begin{figure}[t]
  \centering
  \includegraphics[width=0.95\linewidth]{ 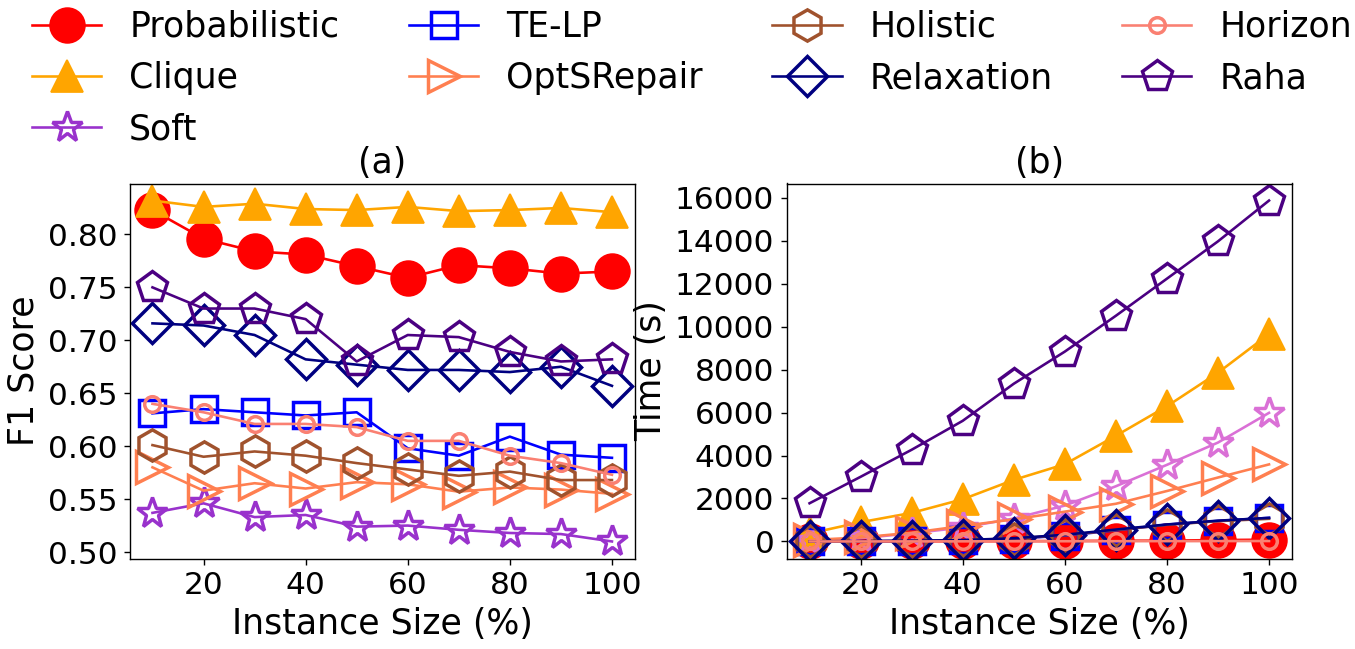}
  \vspace{-0.1in}
  \caption{\textcolor{black}{Performance on Income with varying instance sizes}}
  \label{fig:income}
  \vspace{-0.15in}
\end{figure}

\begin{figure}[t]
  \centering
  \includegraphics[width=1\linewidth]{ 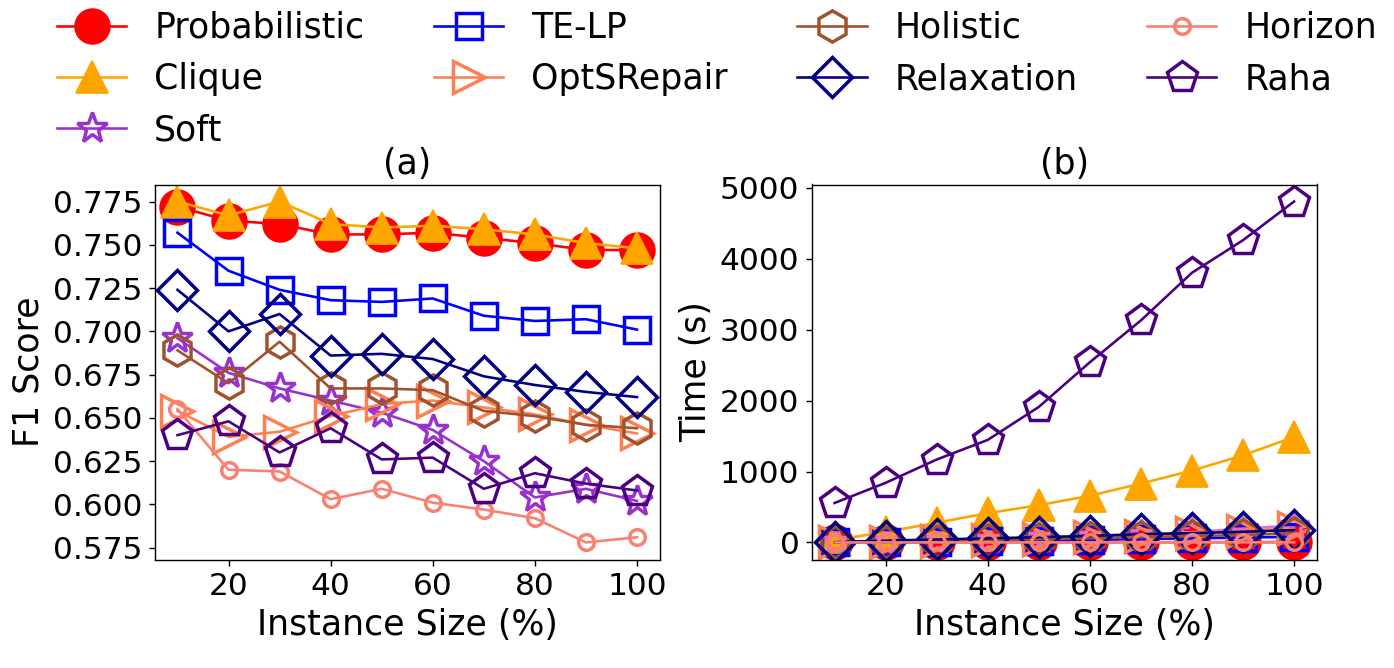}
  \vspace{-0.2in}
  \caption{\textcolor{black}{Performance on Inspection with varying instance sizes}}
  \label{fig:inspection}
  \vspace{-0.2in}
\end{figure}

Figure \ref{fig:income} and \textcolor{black}{Figure \ref{fig:inspection}} show the performance of various methods on the Income (1M) and Inspection (200k) datasets, with instance sizes ranging from $10\%$ to $100\%$, to evaluate the scalability of our methods.
The accuracy of all methods exhibits a slight decline as the instance size increases. \textcolor{black}{A larger instance size may involve more clean tuples in conflicts, thereby complicating the resolution process.}
Among the methods tested, our Clique and Probabilistic approaches achieve similar accuracy and outperform the remaining methods. 
In terms of speed, machine learning-based methods, such as Raha, incur higher time costs due to their complex model structure.
Clique is slower than most of the constraint-based methods as a result of the high complexity associated with linear programming and its iterative framework. Probabilistic, on the other hand, offers comparable efficiency to the other constraint-based methods. With instance size increasing, both Clique and Probabilistic show a mild slowdown, indicating the good scalability of our methods.

\subsection{Evaluating Our Methods}
\label{sect-exp-proposed}



To start, we documented the maximum number of iterations required for Clique on each dataset in Table \ref{tab:Number of Iterations of Clique on each Dataset}.
Across all datasets, Clique terminated in no more than 5 iterations across all datasets, far fewer than the theoretical $C_c^3$. This discrepancy stems from the worst-case analysis, which assumes only one clique of size three per iteration, a rare occurrence.

\begin{table}[h]
\centering
\caption{Maximum Iterations }
\vspace{-0.05in}
\label{tab:Number of Iterations of Clique on each Dataset}
\begin{tabular}{l|c|c|c|c}
\hline
 Dataset & Flights & Rayyan & Restaurant & Soccer   \\
\hline
 Iteration & 3 & 2 & 4 & 2   \\
\hline
 Dataset & Income & Inspection & Iris & Yeast    \\
\hline
 Iteration & 2 & 3 & 1 & 1    \\
\hline
\end{tabular}
\vspace{-0.05in}
\end{table}

\begin{figure}[t]

  \centering
  \includegraphics[width=\linewidth]{ 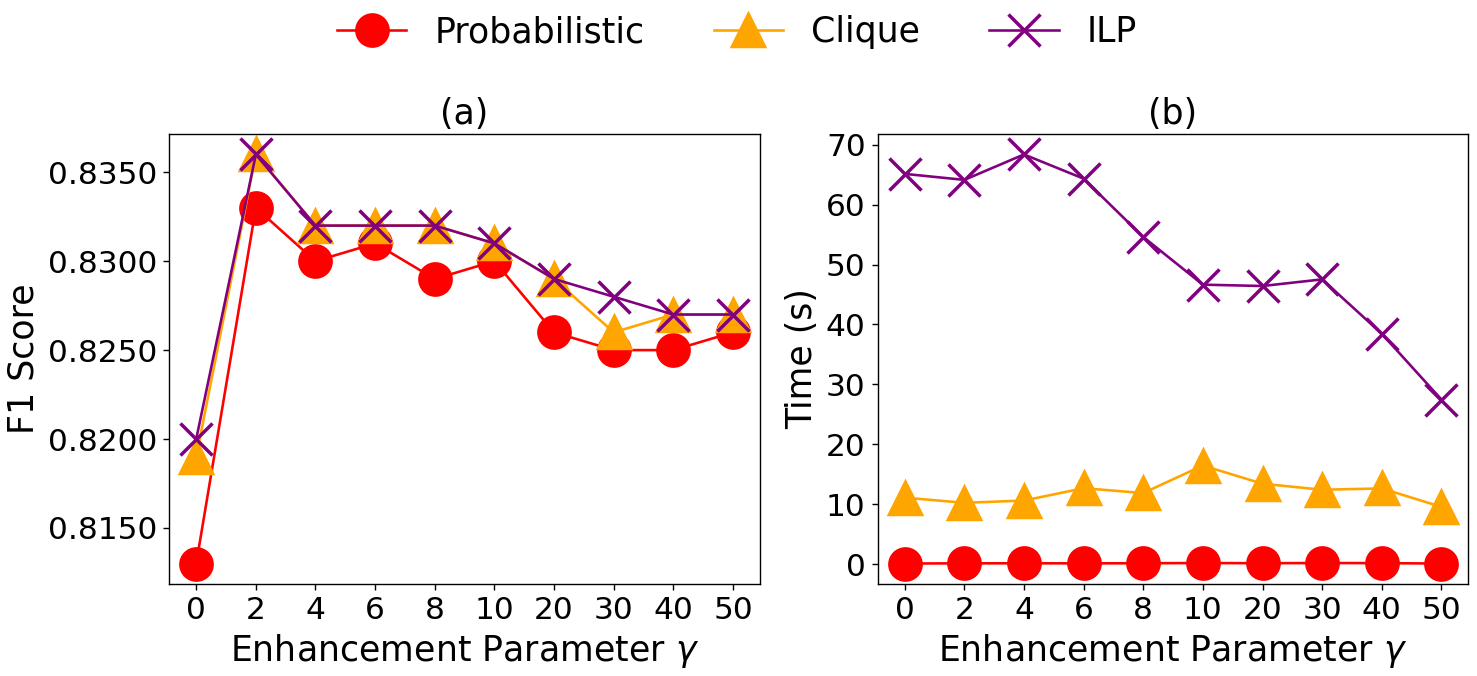}
  \vspace{-0.25in}
  \caption{\textcolor{black}{Performance on various enhancement parameters $\gamma$ over Rayyan}}
    \vspace{-0.15in}
  \label{fig:gamma}
\end{figure}

We analyze the parameter sensitivity of our methods. ILP refers to the exact method in Section \ref{sec: exact method}. Figure \ref{fig:gamma} shows the performance of Probabilistic, Clique, and ILP as $\gamma$ varies from $0$ to $50$. For $\gamma$ between $0$ and $2$, accuracy improves for all methods due to the increased conflict difference. Beyond $\gamma = 2$, accuracy slightly declines as higher $\gamma$ amplifies local information, leading to more locally optimal solutions. In terms of time cost, Clique and Probabilistic show stable performance, while ILP's cost decreases, as higher $\gamma$ helps find exact solutions more easily.

\begin{figure}[t]
  \centering
  \includegraphics[width=\linewidth]{ 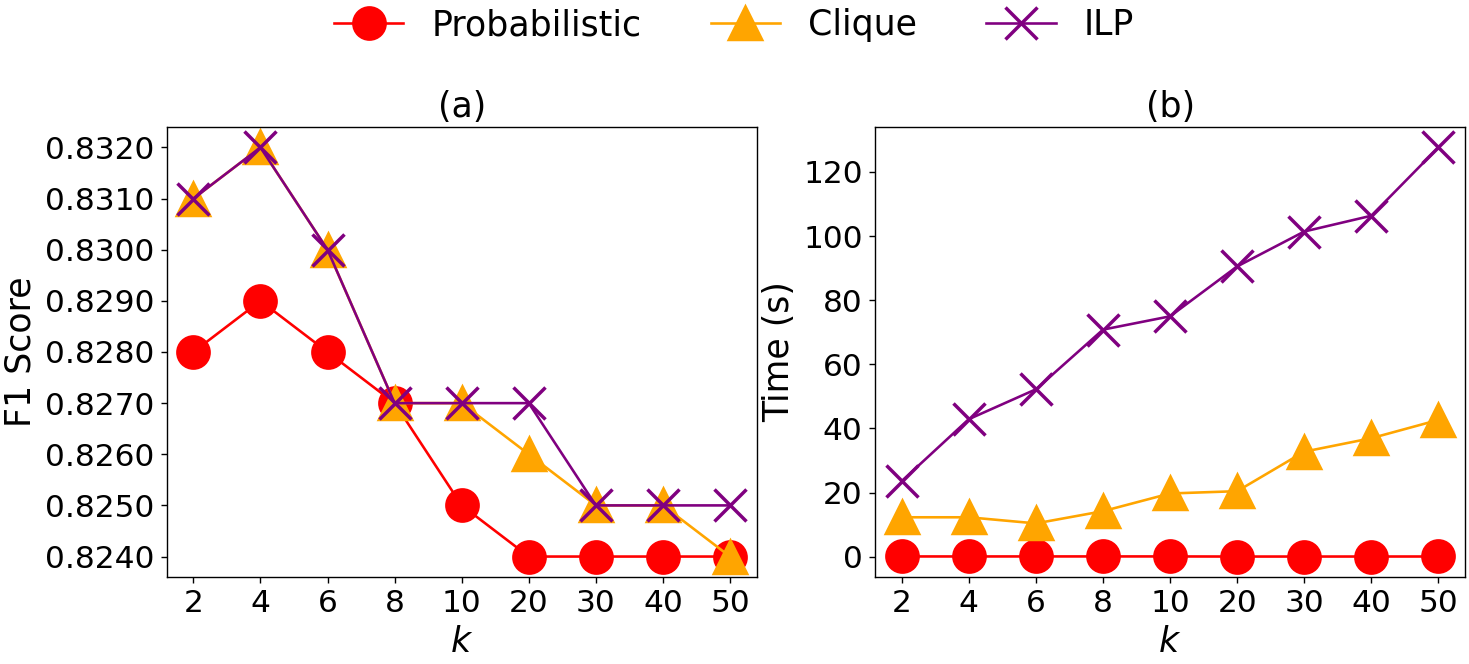}
  \vspace{-0.25in}
  \caption{Performance on various numbers $k$ of $L(t_i,t_l)$ providers for calculating $L(t_i)$ (in Formula \ref{eq: Loss of the remaining set}) over Rayyan}
  \vspace{-0.15in}
  \label{fig:kL}
\end{figure}

\begin{figure}[ht]
  \centering
  \includegraphics[width=\linewidth]{ 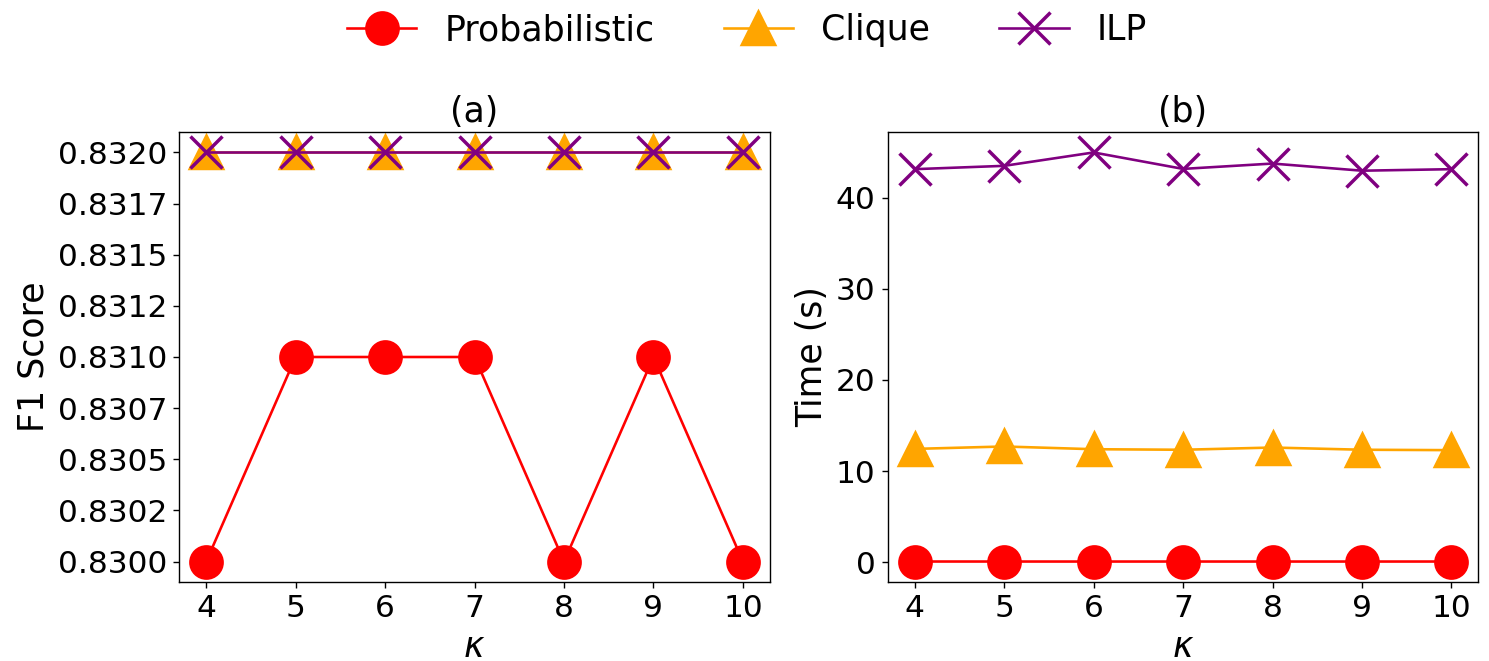} 
  \vspace{-0.25in}
  \caption{Performance on various training sizes $\kappa$ over Rayyan}
  \label{fig:kT}
  \vspace{-0.15in}
\end{figure}

Figure \ref{fig:kL} shows the effect of different $k$ values on method performance. The $F1$-score for all methods increases up to $k=4$, then declines. With a small set of providers, attribute dependency is insufficient, and with too many tuples, differing dependencies harm performance. Regarding time cost, Probabilistic remains stable, while Clique and ILP slow down as $k$ increases due to more $y_{il}$ variables. Figure \ref{fig:kT} shows that both accuracy and time cost remain stable with respect to $\kappa$.

\begin{figure}[ht]
  \centering
  \includegraphics[width=\linewidth]{ 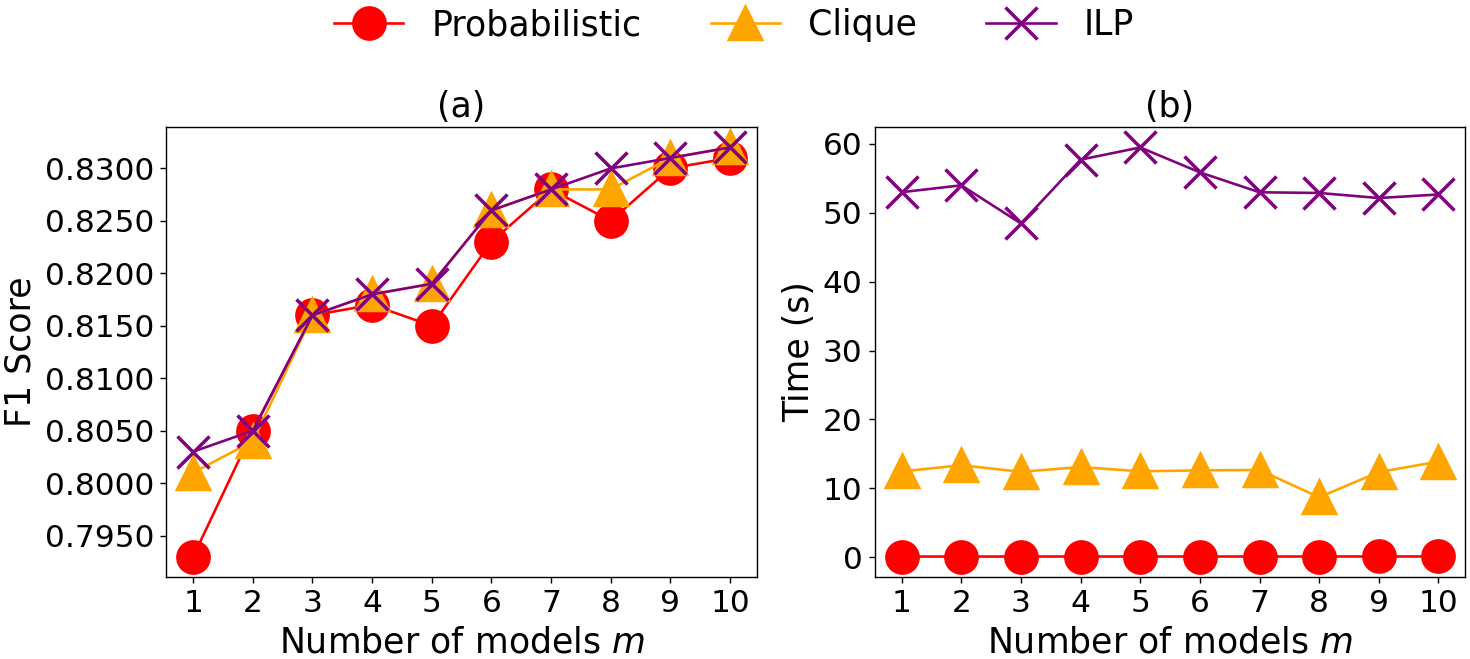} 
  \vspace{-0.25in}
  \caption{Performance on various model amount $m$ over Rayyan}
  \label{fig:Number of models m}
  \vspace{-0.15in}
\end{figure}

\begin{figure}[ht]
  \centering
  \includegraphics[width=\linewidth]{ 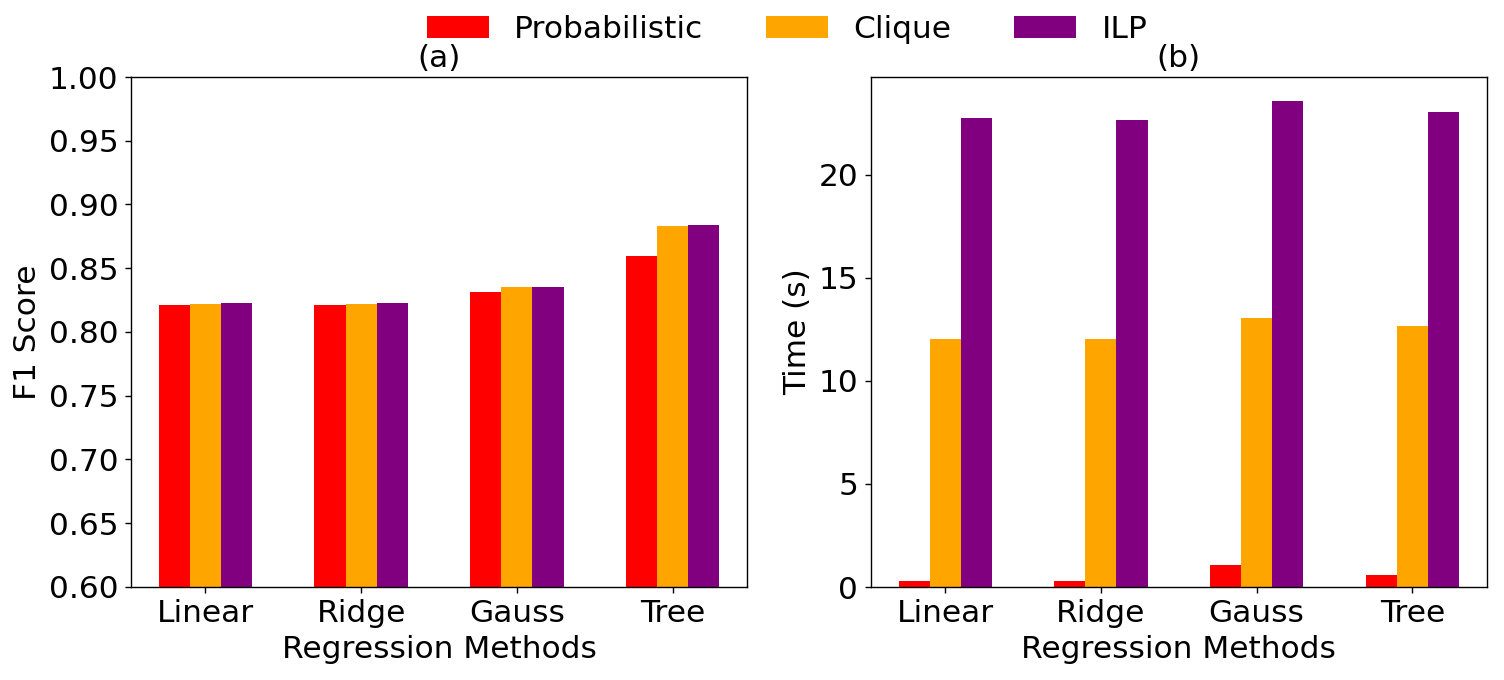} 
  \vspace{-0.25in}
  \caption{\textcolor{black}{Performance on regression models over Rayyan}}
  \label{fig:diff reg models}
  \vspace{-0.15in}
\end{figure}

Figure \ref{fig:Number of models m} shows the effect of varying the number of models $m$ trained for each tuple $t_i$ in dataset $I$. As $m$ increases from 1 to 10, accuracy improves, suggesting that more models capture attribute dependencies better. In Figure \ref{fig:Number of models m} (b), the time cost remains stable across methods. Notably, Clique and ILP yield similar results in most cases, highlighting the optimality of Clique. 
\textcolor{black}{Figure \ref{fig:diff reg models} shows performance with different regression models (Linear, Ridge, Gaussian, and Decision Tree). As model capability increases, repair ability improves, but more time is spent on training and prediction.
}

\subsection{Application Study}

To investigate the efficacy of our method in enhancing downstream classification tasks, we conducted experiments on the Iris and Yeast, and selected the Multilayer Perceptron (MLP) as the classifier. We designated the removal set and the remaining set as the training and testing datasets, respectively. As indicated in Table \ref{tab:MLP application}, when both the training and testing sets consist of $I\backslash I_N$, all methods exhibit the highest accuracy across the four training and testing set configurations, suggesting that classifying on a clean dataset can improve accuracy. In all tested scenarios, our proposed methods, Clique and Probabilistic, yield the highest accuracy, indicating their superior contribution to enhancing downstream applications.




\begin{table}[h]
\centering
\setlength{\tabcolsep}{1.5pt}
\caption{\textcolor{black}{Classification application performance}}
\setlength\tabcolsep{2.3pt}
\resizebox{8.5cm}{!}{
\begin{tabular}{c|cccc||cccc}
\hline
& \multicolumn{4}{c||}{$Iris$} & \multicolumn{4}{c}{$Yeast$} \\
\hline
Train & $I_N$ & $I\backslash I_N$ & $I_N$ & $I\backslash I_N$ & $I_N$ & $I\backslash I_N$ & $I_N$ & $I\backslash I_N$  \\
\hline
Test & $I\backslash I_N$ & $I_N$ & $I_N$ & $ I\backslash I_N$ & $I\backslash I_N$ & $I_N$ & $I_N$ & $I\backslash I_N$  \\
\hline
& \multicolumn{4}{c||}{$5\%$} & \multicolumn{4}{c}{$5\%$} \\
\hline
TE-LP & 0.737 & 0.625 & 0.632 & 0.852 & 0.364 & 0.381 & 0.308 & 0.509 \\
OptSRepair & 0.729 & 0.679 & 0.616 & 0.849 & 0.412 & 0.384 & 0.242 & 0.485 \\
Soft & 0.694 & 0.759 & 0.656 & 0.796 & 0.426 & 0.309 & 0.261 & 0.505 \\
Relaxation  & 0.748 & 0.718 & 0.680 & 0.806 & 0.396 & 0.300 & 0.268 & 0.511 \\
Holistic & 0.756 & 0.643 & 0.632 & 0.840 & 0.332 & 0.352 & 0.244 & 0.492 \\
Horizon & 0.669 & 0.760 & 0.678 & 0.740 & 0.451 & 0.311 & 0.440 & 0.506 \\
Raha & 0.400 & 0.603 & 0.280 & 0.860 & 0.386 & 0.299 & 0.331 & 0.510 \\
HC & 0.720 & 0.320 & 0.453 & 0.849 & 0.443 & 0.326 & 0.333 & 0.483 \\
Probabilistic & 0.741 & 0.612 & 0.686 & 0.872 & 0.331 & 0.296 & 0.414 & 0.514 \\
Clique & 0.713 & 0.648 & 0.620 & \textbf{0.883} & 0.453 & 0.278 & 0.333 & \textbf{0.529} \\
\hline
& \multicolumn{4}{c||}{$10\%$} & \multicolumn{4}{c}{$10\%$} \\
\hline
TE-LP & 0.534 & 0.628 & 0.399 & 0.704 & 0.364 & 0.305 & 0.273 & 0.439 \\
OptSRepair & 0.578 & 0.668 & 0.508 & 0.746 & 0.405 & 0.269 & 0.325 & 0.444 \\
Soft & 0.705 & 0.647 & 0.542 & 0.741 & 0.393 & 0.321 & 0.375 & 0.445 \\
Relaxation & 0.700 & 0.682 & 0.513 & 0.778 & 0.396 & 0.310 & 0.255 & 0.438 \\
Holistic & 0.560 & 0.613 & 0.393 & 0.698 &  0.348 & 0.292 & 0.251 & 0.433 \\
Horizon & 0.628 & 0.509 & 0.648 & 0.719 &  0.389 & 0.305 & 0.389 & 0.448 \\
Raha & - & - & - & 0.754 & 0.364 & 0.313 & 0.335 & 0.437 \\
HC & 0.740 & 0.439 & 0.360 & 0.775 &  0.372 & 0.372 & 0.340 & 0.424 \\
Probabilistic & 0.685 & 0.647 & 0.650 & 0.803 & 0.374 & 0.306 & 0.309 & 0.452 \\
Clique & 0.740 & 0.692 & 0.670 & \textbf{0.855} & 0.367 & 0.319 & 0.339 & \textbf{0.457} \\
\hline
& \multicolumn{4}{c||}{$15\%$} & \multicolumn{4}{c}{$15\%$} \\
\hline
TE-LP & 0.663 & 0.522 & 0.617 & 0.721 & 0.359 & 0.302 & 0.296 & 0.392 \\
OptSRepair & 0.599 & 0.508 & 0.524 & 0.691 & 0.368 & 0.296 & 0.289 & 0.397 \\
Soft & 0.621 & 0.582 & 0.493 & 0.742 & 0.365 & 0.289 & 0.342 & 0.403 \\
Relaxation &0.767 & 0.467 & 0.540 & 0.728 & 0.333 & 0.309 & 0.292 & 0.400 \\
Holistic &  0.673 & 0.514 & 0.627 & 0.718 & 0.385 & 0.314 & 0.398 & 0.406 \\
Horizon &  0.510 & 0.569 & 0.550 & 0.716 & 0.375 & 0.276 & 0.350 & 0.404 \\
Raha &  0.660 & 0.597 & 0.540 & 0.750 &  0.304 & 0.286 & 0.295 & 0.400 \\
HC &  0.416 & 0.424 & 0.432 & 0.736 &  0.359 & 0.284 & 0.280 & 0.395 \\
Probabilistic &   0.739 & 0.505 & 0.720 & 0.752 & 0.348 & 0.303 & 0.337 & 0.409 \\
Clique &  0.617 & 0.535 & 0.548 & \textbf{0.768} & 0.384 & 0.318 & 0.344 & \textbf{0.420} \\
\hline
& \multicolumn{4}{c||}{$20\%$} & \multicolumn{4}{c}{$20\%$} \\
\hline
TE-LP & 0.708 & 0.634 & 0.602 & 0.686 & 0.325 & 0.303 & 0.284 & 0.391 \\
OptSRepair & 0.747 & 0.692 & 0.613 & 0.721 & 0.298 & 0.318 & 0.288 & 0.382 \\
Soft & 0.688 & 0.657 & 0.584 & 0.696 & 0.342 & 0.267 & 0.329 & 0.374 \\
Relaxation &  0.633 & 0.544 & 0.608 & 0.704 &  0.316 & 0.333 & 0.264 & 0.380 \\
Holistic &  0.720 & 0.653 & 0.573 & 0.656 &  0.311 & 0.325 & 0.246 & 0.383 \\
Horizon  & 0.703 & 0.654 & 0.646 & 0.686 &  0.359 & 0.287 & 0.338 & 0.377 \\
Raha &  0.748 & 0.422 & 0.440 & 0.769 & 0.368 & 0.323 & 0.386 & 0.386 \\
HC &  0.706 & 0.772 & 0.634 & 0.738 & 0.340 & 0.374 & 0.378 & 0.354 \\
Probabilistic &  0.537 & 0.610 & 0.714 & 0.772 & 0.352 & 0.327 & 0.346 & 0.403 \\
Clique &  0.708 & 0.587 & 0.737 & \textbf{0.788} &  0.389 & 0.378 & 0.347 & \textbf{0.404} \\
\hline
\end{tabular}
}
\vspace{-0.15in}
\label{tab:MLP application}
\end{table}

\section{Related Work}
\label{sect-related}


In this section, we discuss the representative data-repairing methods relying on constraints, statistics, neighborhood, and machine learning techniques.


\textbf{Constraint-based} methods harness a particular class of integrity constraints, such as functional dependencies (FDs) \cite{DBLP:conf/kdd/MandrosBV17}, conditional functional dependencies (CFDs) \cite{DBLP:journals/tods/FanGJK08}, matching dependencies (MDs) \cite{DBLP:conf/icdt/BertossiKL11}, and denial constraints (DCs) \cite{DBLP:conf/icde/ChuIP13}, to maintain data consistency. The majority of these methods adhere to the principle of minimal repair \cite{DBLP:conf/icde/ChuIP13,DBLP:conf/sigmod/SongZW16}. \textcolor{black}{OptSRepair \cite{DBLP:conf/pods/LivshitsKR18} is a method for FD conflicts. With value frequency and maximum matching, the method could get a theoretically optimal solution in certain cases}. Holistic \cite{DBLP:conf/icde/ChuIP13} is a constraint-based approach that considers the interplay between various constraints. It transforms conflicts into a hypergraph structure and identifies a minimum vertex cover as the error cells. 
Horizon \cite{DBLP:journals/pvldb/RezigOAEMS21} is a constraint-based method tailored for FDs. It leverages the most prevalent patterns for repairs and attains linear computational complexity.

\textbf{Statistics-based} cleaning approaches mainly leverage statistical information to repair dirty values. ERACER \cite{DBLP:conf/sigmod/MayfieldNP10} uses belief propagation and relational dependency networks for the data-repairing task, offering the advantage of handling complex problems but requiring more computational resources. SCARE \cite{DBLP:conf/sigmod/YakoutBE13} employs a simpler naive Bayesian model, resulting in potentially higher efficiency but limited capability to handle certain complex issues. These methods have been shown  less effective by  existing works \cite{DBLP:journals/pvldb/RekatsinasCIR17,DBLP:conf/sigmod/YakoutBE13}, so we did not conduct experiments on these algorithms.

\textbf{Neighborhood-based} methods leverage clustering or density to rectify errors. DORC \cite{DBLP:conf/kdd/SongLZ15} integrates both data cleaning and clustering techniques. It identifies tuples with low density as outliers and relocates them to the neighborhoods of core points. Relaxation \cite{DBLP:journals/tkde/SunSY24} employs integrity constraints to detect errors under the guidance of density. 
This method overcomes the limitation of being unable to select among multiple potential removal sets
Nonetheless,
as shown in Example \ref{eg:example-motivation2}, tuples with higher frequency may also harbor errors. This oversight arises from a failure to scrutinize the attribute values within the tuples, an issue that our work aims to address.

\textbf{ML-based} techniques rely on models to capture complex information. Raha \cite{DBLP:conf/sigmod/MahdaviAFMOS019} and Baran \cite{DBLP:journals/pvldb/MahdaviA20} are a pair of methods in which the former is used for error detection and the latter fixes them. Raha  
is a  configuration-free system. It
combines various error detection techniques into one framework to detect the errors. Baran  is an error correction system that combines context awareness and transfer learning for repairing. 
Holoclean \cite{DBLP:journals/pvldb/RekatsinasCIR17} integrates diverse signals like integrity constraints, external data sources, and dataset-specific statistics to generate a probabilistic model for value inference. Such methodologies harness more complex signals, which, while enhancing their capabilities, also increase their complexity.

\section{Conclusion}
In this work, considering the limitations of existing approaches that solely depend on frequent values, we identify the optimal S-repair with respect to denial constraints by leveraging attribute dependencies. To guarantee that the remaining tuple set adheres to constraints while conforming to the dependencies, we (1) formalize the problem into an Integer Linear Program (ILP) for the exact solution in Section \ref{sec: exact method}; (2) introduce a clique-based algorithm that offers approximate guarantees and certain case optimality in Section \ref{sec:LP-relax}; and (3) develop a probabilistic approach with a defined error margin in Section \ref{sec:prob based method}. Comprehensive experiments on datasets with real-world and synthetic errors validate the efficacy of our methods in S-repair, 
and subsequent tasks.

\clearpage
\section*{\textcolor{black}{AI-Generated Content Acknowledgement}}
\textcolor{black}{This paper was initially written by the authors, and certain sections (including the Introduction, section \ref{sec:exp}, \ref{sect-related}) of the text were subsequently refined using the ChatGPT-4 language model (OpenAI, 2023). The AI-generated content was reviewed, edited, and approved by the authors to ensure accuracy and consistency with the research objectives. All final revisions were made manually by the authors to ensure the clarity and correctness of the language.}

\bibliographystyle{IEEEtran}
\bibliography{ref}

\vspace{12pt}

\end{document}